\documentclass[a4paper,11pt]{article}
\usepackage{jinstpub} 
\usepackage{lineno}
\usepackage{comment}

\begin{document}

\title{Light exposure and temperature effects on quantum efficiency of bialkali metal photomultipier tubes}
\author[a,b]{Antonio~De~Benedittis}\note{Corresponding author. e-mail: antonio.debenedittis@na.infn.it}
\author[b]{Pasquale~Migliozzi}
\author[b]{Carlos~Maximiliano~Mollo}
\author[b]{Andreino~Simonelli}
\author[a,b]{Daniele~Vivolo}
\affiliation[a]{Universit{\`a} degli Studi della Campania "Luigi Vanvitelli", Dipartimento di Matematica e Fisica, viale Lincoln 5, Caserta, 81100 Italy}
\affiliation[b]{INFN, Sezione di Napoli, Complesso Universitario di Monte S. Angelo, Via Cintia ed. G, Napoli, 80126 Italy}

\abstract{In astroparticle experiments, photomultiplier tubes are crucial for detecting Cherenkov radiation emitted by charged particles, owing to their exceptional sensitivity to low-intensity light, which is essential for studying high-energy phenomena associated with astrophysical neutrinos. However, due to their high sensitivity, PMTs are vulnerable to significant damage to their photocathode coating when exposed to intense light and/or high temperatures. Although this scenario is rare under controlled conditions, it can become relevant in experiments with prolonged exposure to sunlight and elevated temperatures.

This study presents an analysis of the damage threshold and recovery time of photomultipliers with bialkali metal coatings. The investigation involved measuring the quantum efficiency of the PMTs before and after exposure to a xenon lamp for varying durations, simulating sunlight exposure over several days. Additionally, quantum efficiency was assessed before and after the PMTs were subjected to thermal stress, providing an evaluation of their performance under different thermal conditions.}

\keywords{Photomultiplier; Quantum Efficiency; Dark Current; Light exposure; Thermal stress}

\maketitle
\flushbottom

\section{Introduction}

In the field of astroparticle physics, the performance of photomultiplier tubes (PMTs) is critically important, especially for experiments that detect Cherenkov radiation \cite{CHER} produced by high-energy cosmic particles. Cherenkov radiation occurs when charged particles exceed the speed of light in a given medium and is a key phenomenon in the study of cosmic rays, gamma rays, and neutrinos \cite{hess, LHAASO, HAWC, magic, IC, KM}.

The Quantum Efficiency (QE) of PMTs, representing the probability of converting photons at specific wavelengths in photoelectrons, is essential for accurately determining the energy and direction of these particles. Precise QE measurements are crucial in various experimental environments. These measurements significantly affect the sensitivity and reliability of Cherenkov detectors and are fundamental for accurate Monte Carlo simulations.

Given the extensive use of PMTs in astroparticle experiments, ensuring their optimal performance through detailed characterisation is essential. The PMT photocathode is the active component that interacts with photons generated by particle interactions in the medium. Thus, understanding the performance characteristics of PMTs, including their QE and dark noise, is crucial. However, PMTs are highly sensitive devices and can suffer damage, particularly to the photocathode coating, which may affect their performance \cite{hamam}. Although such damage is rare in controlled laboratory environments, it is possible during some experimental phases, in which PMTs are exposed to intense light or thermal sources.


Several studies have investigated the degradation mechanisms affecting the QE of PMTs and MCP-PMTs under various conditions \cite{gadza, jinno}. However, systematic studies on the reversibility of QE degradation have been primarily conducted in the context of OLEDs and metallic photocathodes \cite{perrone, seon, yahiro}, while such investigations remain rare in the field of photomultiplier tubes.

This study investigates the QE and dark current behaviour of Hamamatsu bialkali Sb-K-Cs metal-coated 3-inch PMTs under light and thermal stress conditions. QE measurements were performed before and after exposing the PMTs to a xenon lamp, simulating prolonged sunlight exposure, and following thermal stress cycles to evaluate performance under varying thermal
conditions. Interestingly, during our investigation, we observed unexpected reversibility patterns in QE and dark current degradation.

The results of this study provide crucial insights into the resilience and performance of PMTs under harsh environmental conditions, ensuring their optimal function in Cherenkov detectors. These findings are significant not only for astroparticle experiments but also for the broader scientific community. By understanding and mitigating the effects of light and thermal stress on PMTs, we can improve the durability and accuracy of these detectors, enhancing their applications in various scientific and industrial fields.

\section{Experimental setup}

\begin{figure}[h!]
    \centering
    \includegraphics[scale=.3]{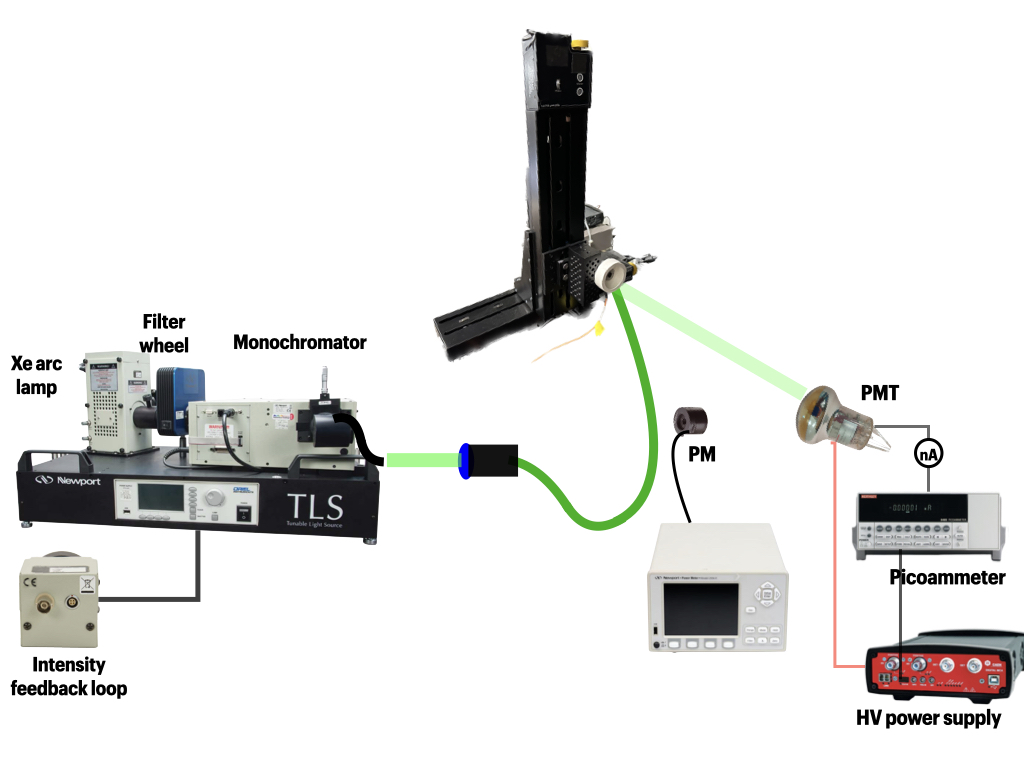}
    \caption{Sketch of the experimental setup with the optical and electrical connections.}
    \label{fig:Setup}
\end{figure}

\figurename~\ref{fig:Setup} (also depicted in \cite{set}) shows the experimental setup used for the measurement of QE across different wavelengths. This is achieved by directing light from a lamp to the center of the PMT and scanning across wavelengths from 280 to 700 nm (that is, the range of wavelengths to which the studied PMTs are sensitive) to map the QE distribution. 
The incident light power is measured using a Newport 918D-UV-OD3R calibrated power probe, connected to and read by a Newport 2936 R power meter. The emission of the NIST-calibrated photodiode has varying uncertainties: 3.4\% in the 220-300 nm range, 1.65\% in the 300-430 nm range, and 1.1\% in the 430-1000 nm range. Current measurement from the first dynode is performed using a Keithley 6485 picoammeter, which offers 0.4\% accuracy in the 20 nA range used for PMT QE assessments. Calibration of these instruments supports all the experimental procedures detailed below.

To avoid underestimating the emitted photocurrent, the photocathode is kept at -$200$ V relative to the first dynode, which is grounded as the other nine dynodes and the PMT anode. This configuration maximizes the capture of photoelectrons, particularly those that might otherwise be missed at the first dynode.

The setup includes a Newport TLS 260 tunable light source, a 300 watt xenon arc lamp with high UV output and excellent spectral performance \cite{lamp}. The light passes through a motorised filter wheel with four low-pass filters to minimise higher-order diffraction, which can lead to stray light and systematic errors. These errors are further reduced by using a Czerny-Turner monochromator, which provides superior spectral purity. The light exiting the monochromator is transmitted through a multimode optical fibre with a 200 µm pure silica core and fluorine-doped silica cladding, designed to minimize stray light.

A light collimator, mounted on a Z-axis stage that is mechanically coupled to an X-axis stage, ensures precise alignment. The LTS300C from Thorlabs provides a versatile system with a 300 mm travel range and a maximum velocity of 50 mm/s. This setup enables precise and rapid movements, with bidirectional repeatability within $\pm$2 $\mu$m and an accuracy of 5 $\mu$m.

\begin{figure}[h]
    \centering
    \includegraphics[scale=.130]{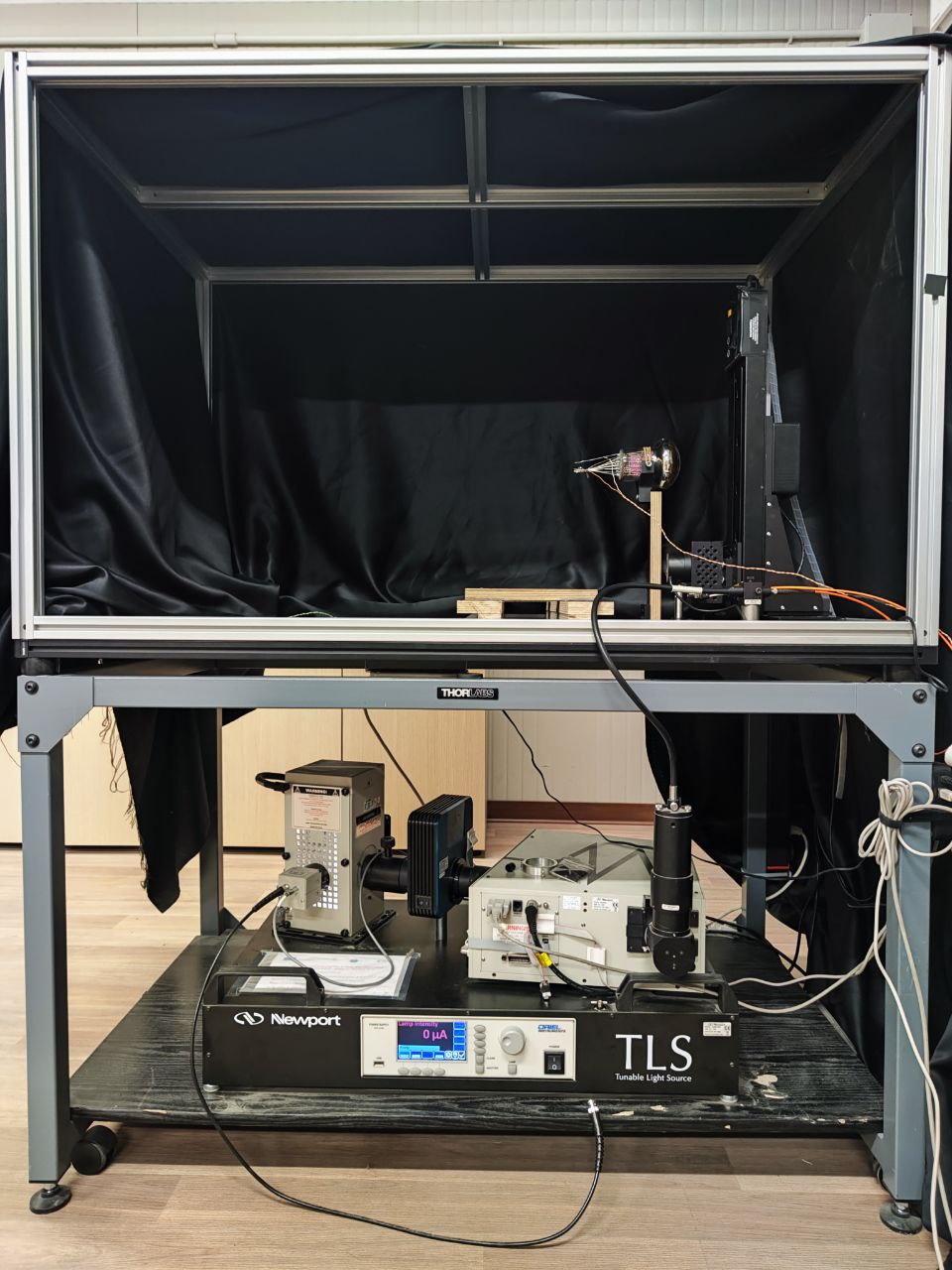}
    \hspace{1cm} 
    \includegraphics[scale=.126]{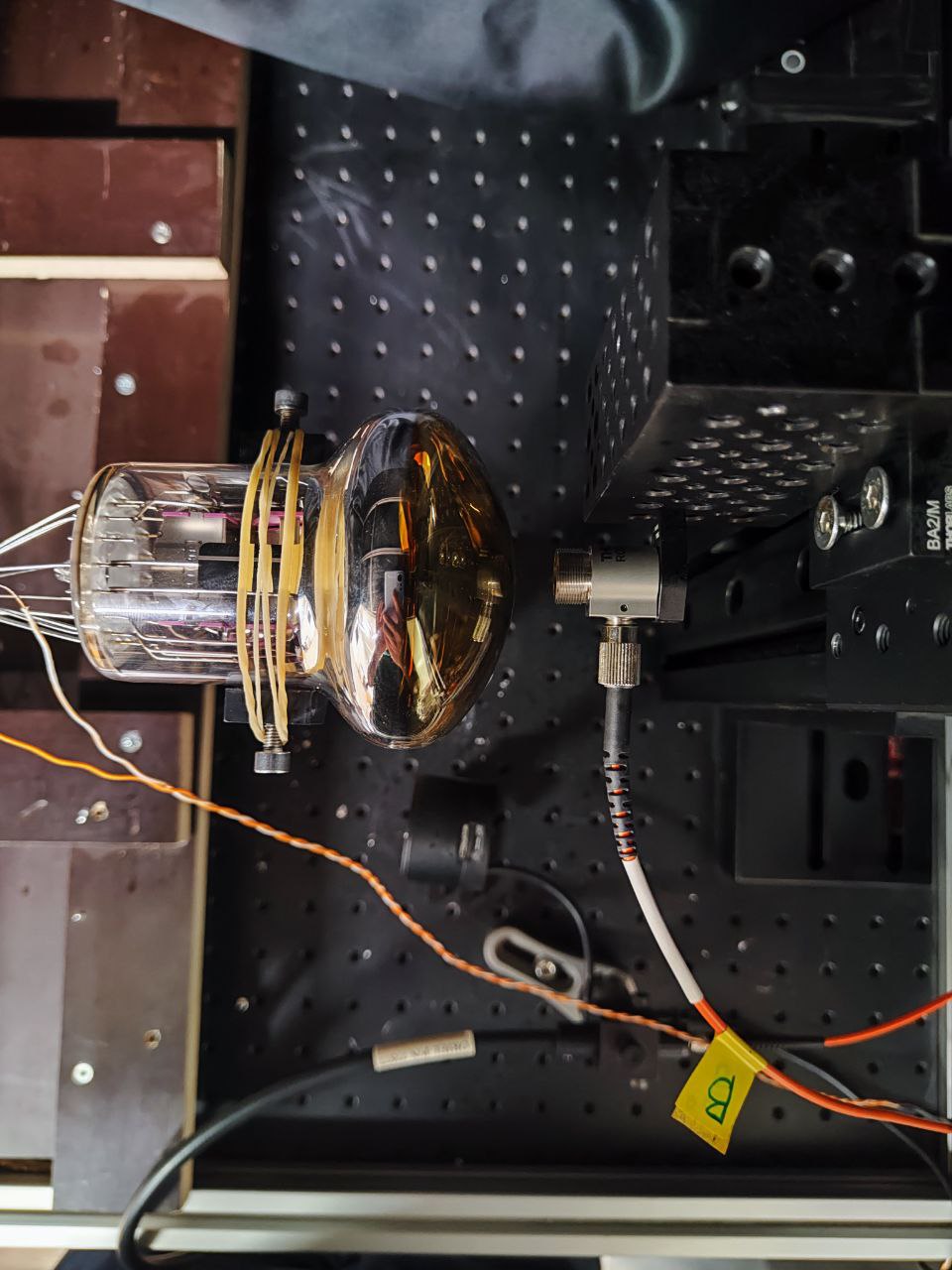}
    \caption{On the left picture, the ensemble of the instruments is displayed. On the right picture, a detail of the optical bench setup shows the device under test and the calibrated power meter.}
    \label{fig:Setup_}
\end{figure}

All components of the setup are shown in \figurename~\ref{fig:Setup_}. The system is controlled via MATLAB’s Instrument Control Toolbox. Communication with peripherals and instruments is managed through RS-232 for the Keithley picoammeter and the monochromator, while the power meter is controlled via USB with a dedicated library \cite{USB}.

The 2D stage operates with custom-developed drivers \cite{THOR} tailored to our requirements, allowing customization in the control code and overall setup. The MATLAB code for QE measurement operates in two phases. First, it averages power and current values over a user-defined number of measurements at a fixed wavelength. In the second phase, this procedure is repeated across the entire wavelength range from 280 to 700 nm in 10 nm increments.

\section{Measurements and results}

Using the experimental setup described in the previous section, we evaluated the QE distribution for several PMTs. To further analyze their performance under stress conditions, we subjected part of these PMTs to light stress. This consisted of exposing them to a 300 watt xenon arc lamp (Figure 3, left), which provided an irradiance of $1.220\ \mathrm{W/cm^2}$, simulating prolonged sun exposure over several days.

\begin{figure}[h]
    \centering
    \includegraphics[scale=.0365]{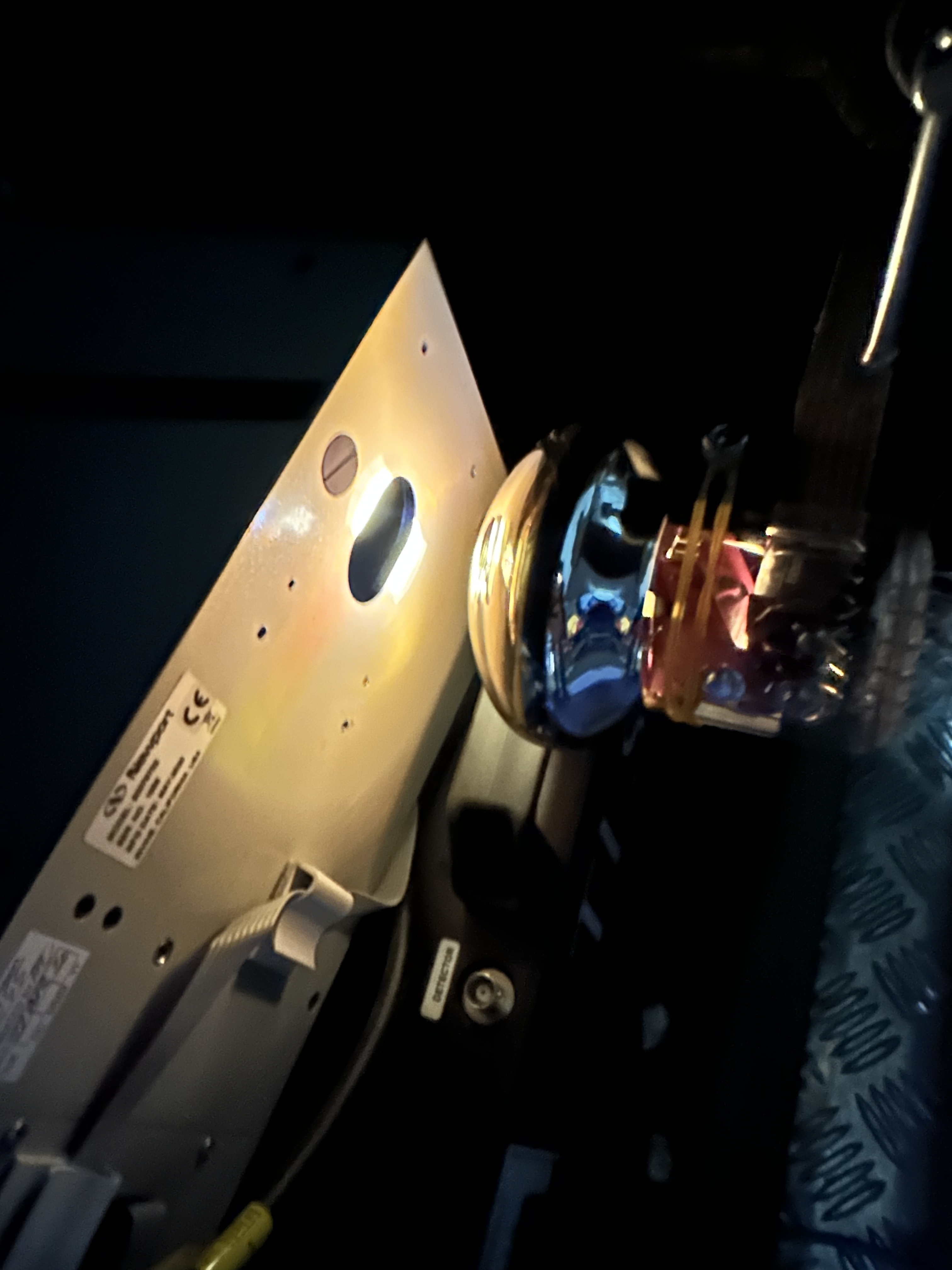}
    \hspace{1cm} 
    \includegraphics[scale=.15]{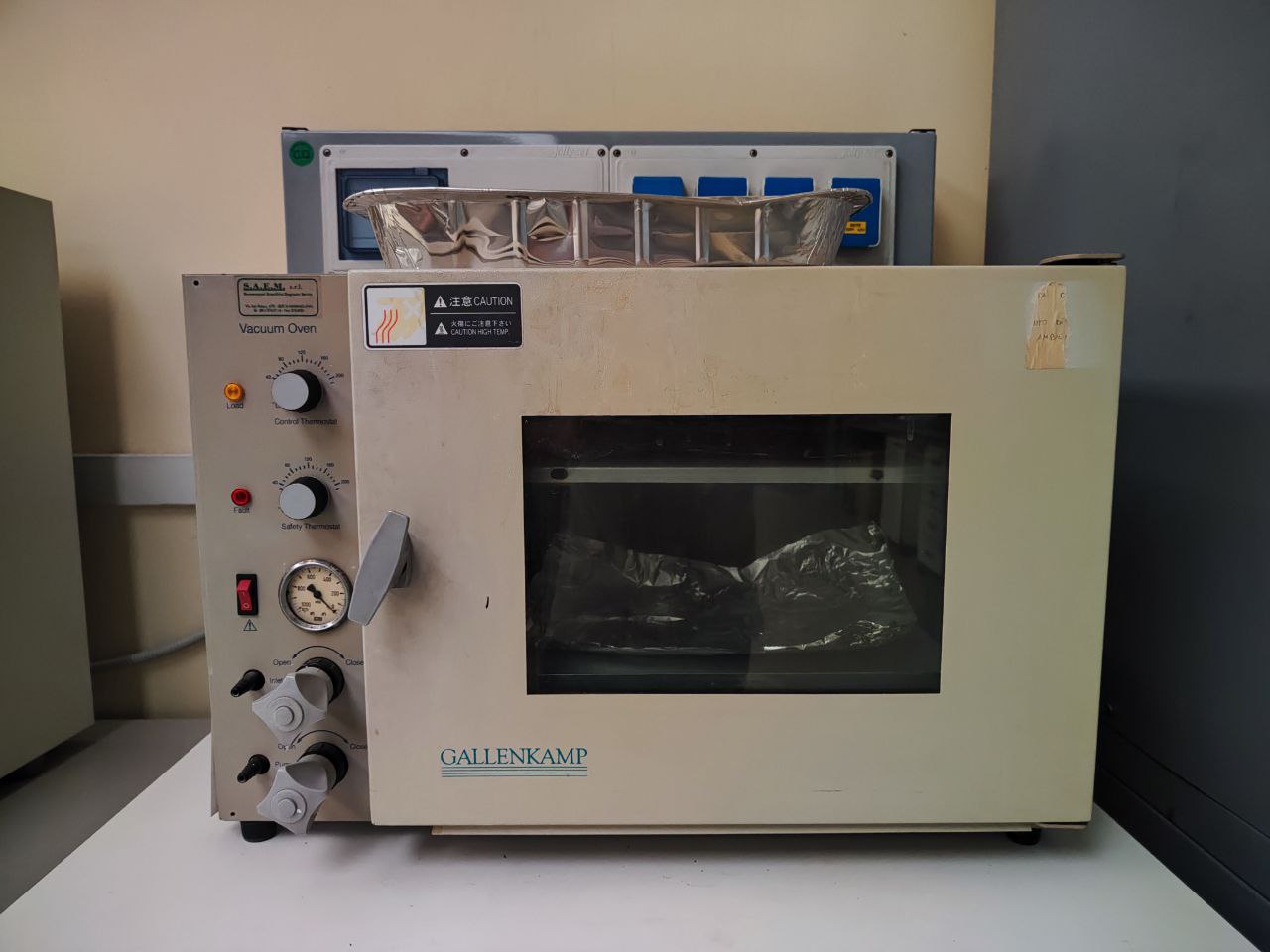}
    \caption{Left: Picture of a PMT exposed to direct light of the 300-watt xenon arc lamp. Right: Gallenkamp vacuum oven used to subject the PMTs to thermal stress.}
    \label{fig:meas}
\end{figure}

Simultaneously, another set of PMTs was subjected to thermal stress. These PMTs were placed in a Gallenkamp vacuum oven (\figurename~\ref{fig:meas}, right), where they were exposed to temperatures ranging from 50°C to 210°C for varying durations. This procedure was designed to assess the impact of extreme temperature variations on the PMTs’ performance.

\subsection{Xenon lamp exposure}

\begin{figure}[h]
    \centering
    \includegraphics[scale=.197]{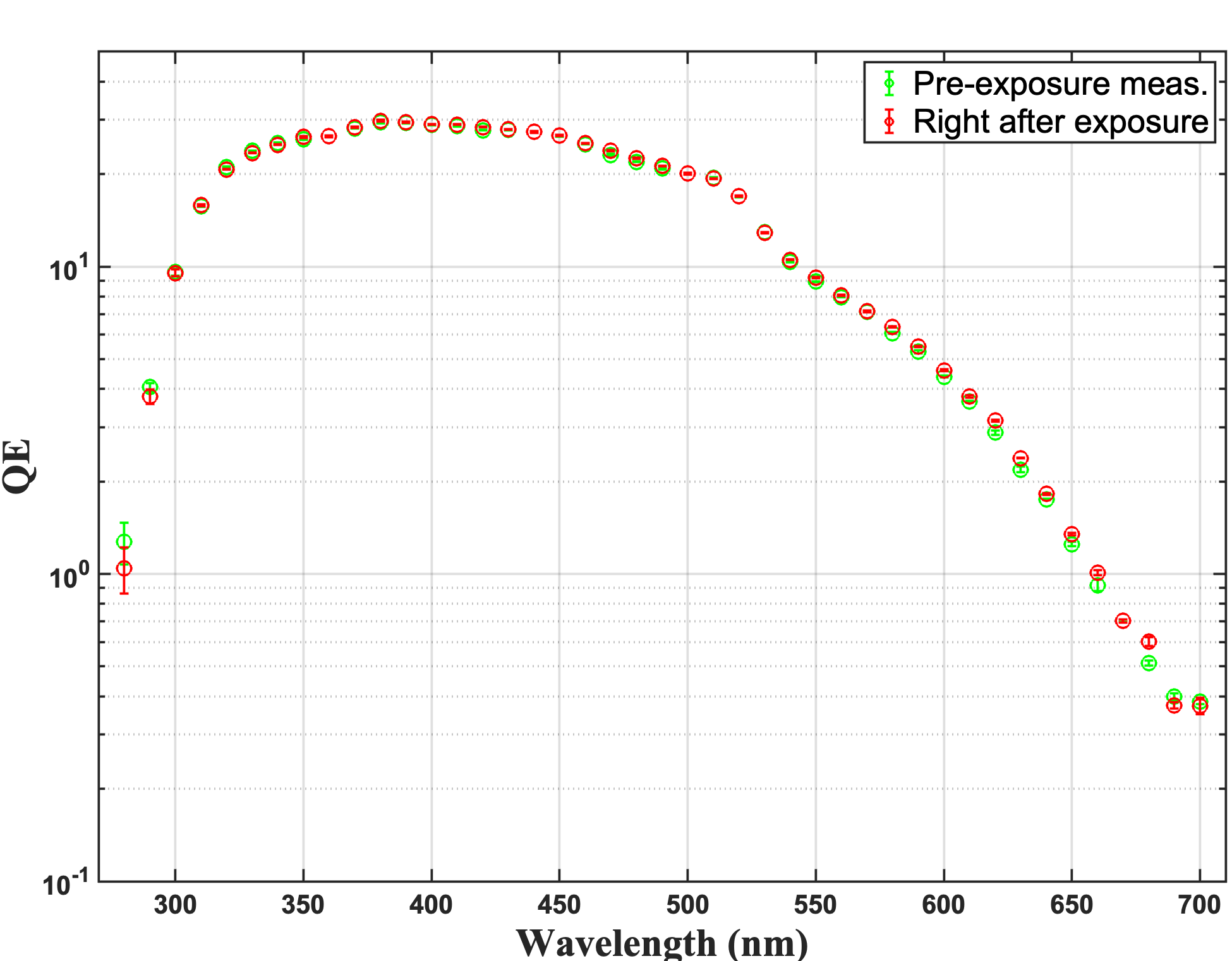}
    \includegraphics[scale=.197]{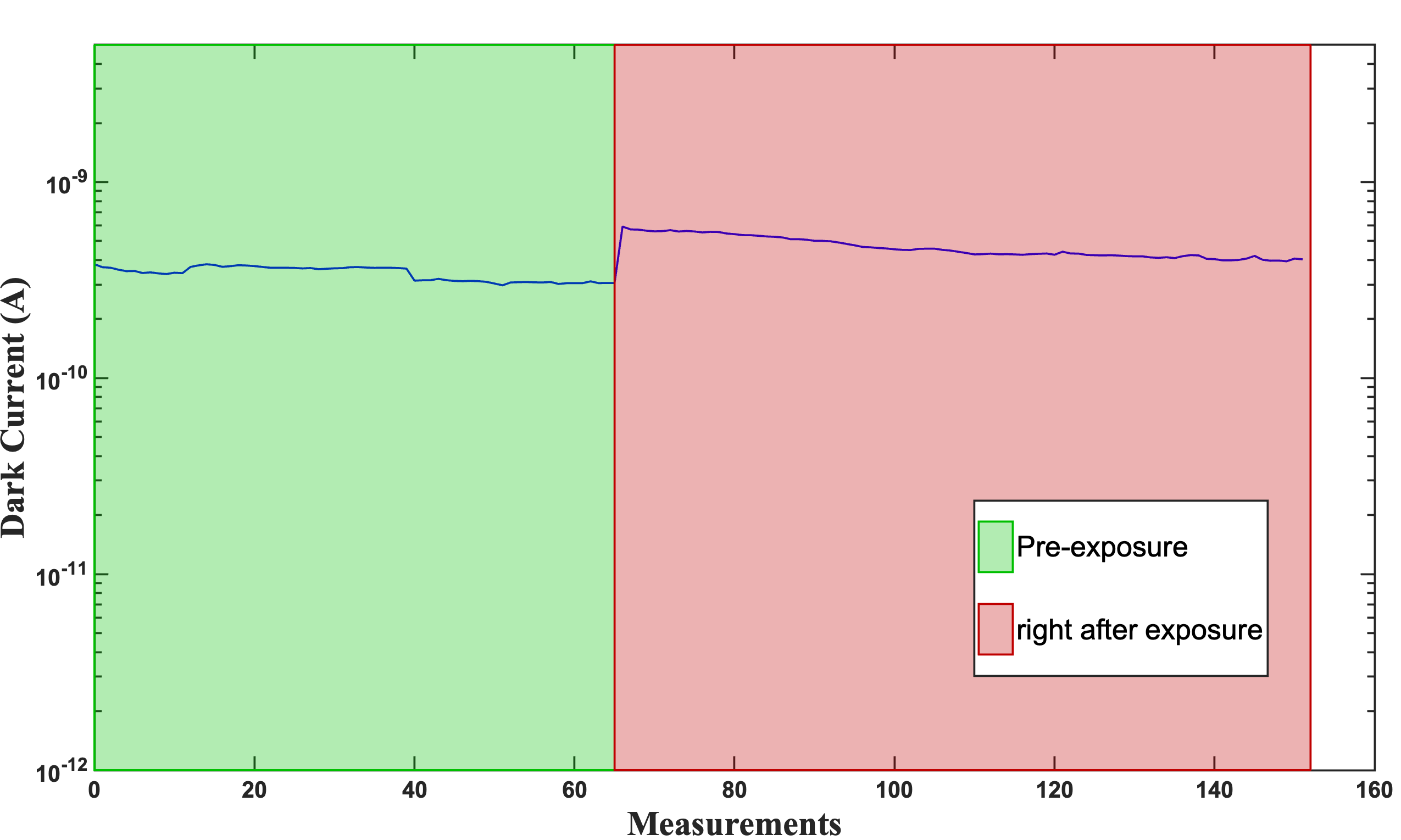}
    \caption{Quantum efficiency as a function of wavelength (left) and the evolution of dark current (right) before and after three hours of lamp exposure.}
    \label{fig:lamp1}
\end{figure}

Three PMTs were exposed to the Xenon lamp light for 3, 23, and 70 hours, simulating sunlight exposure for about 5, 40, and 120 days, respectively, based on a solar irradiance of 85.35 mW/cm$^2$ at a wavelength of 555 nm. In \figurename~\ref{fig:lamp1}, the left side shows the QE trend for a PMT exposed for three hours, with green and red dots representing pre- and post-exposure data, respectively. The right side of the figure illustrates the temporal evolution of the dark current. The data show a good agreement between the pre- and post-exposure trends.

\begin{figure}[h]
    \centering
    \includegraphics[scale=.197]{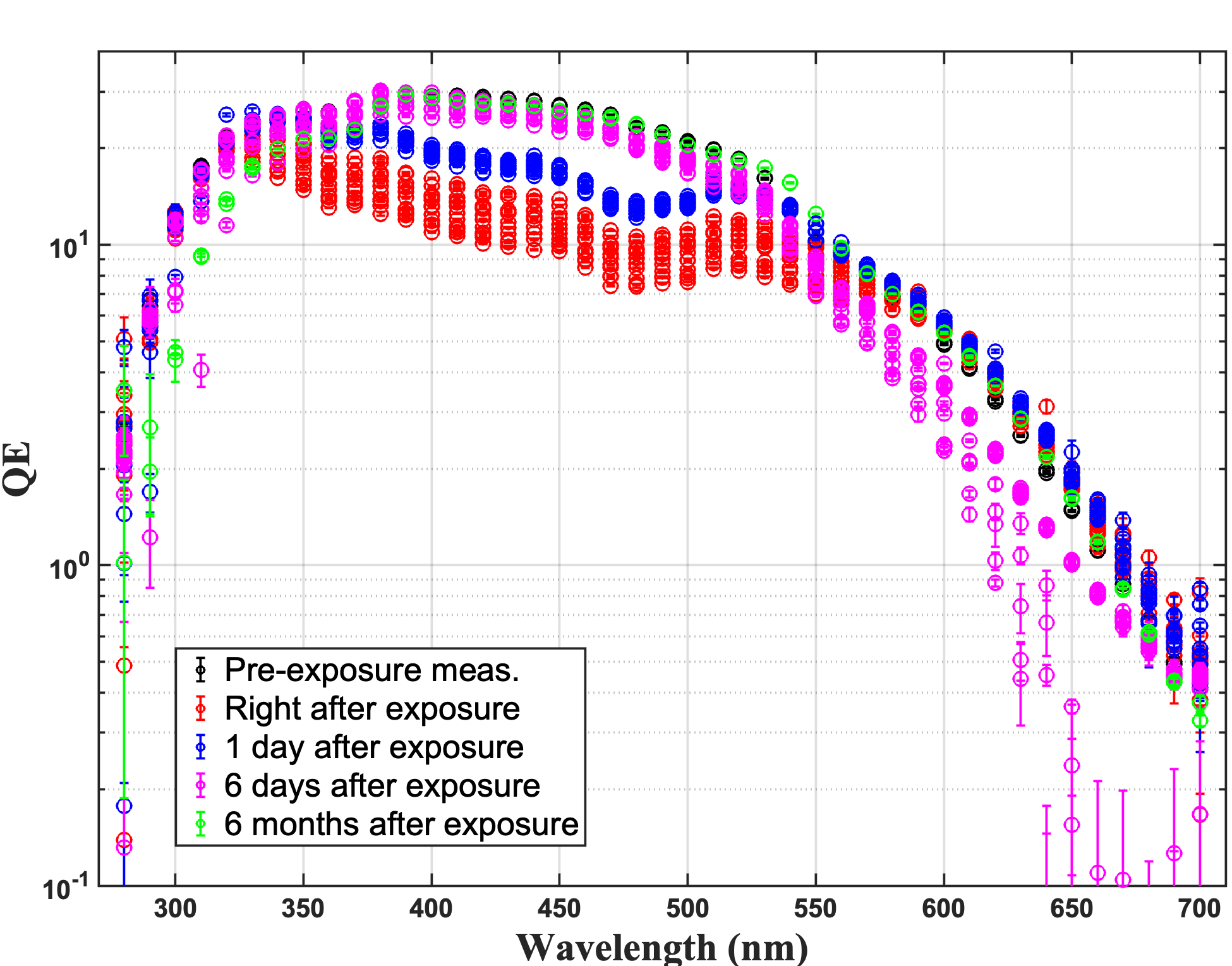}
    \includegraphics[scale=.197]{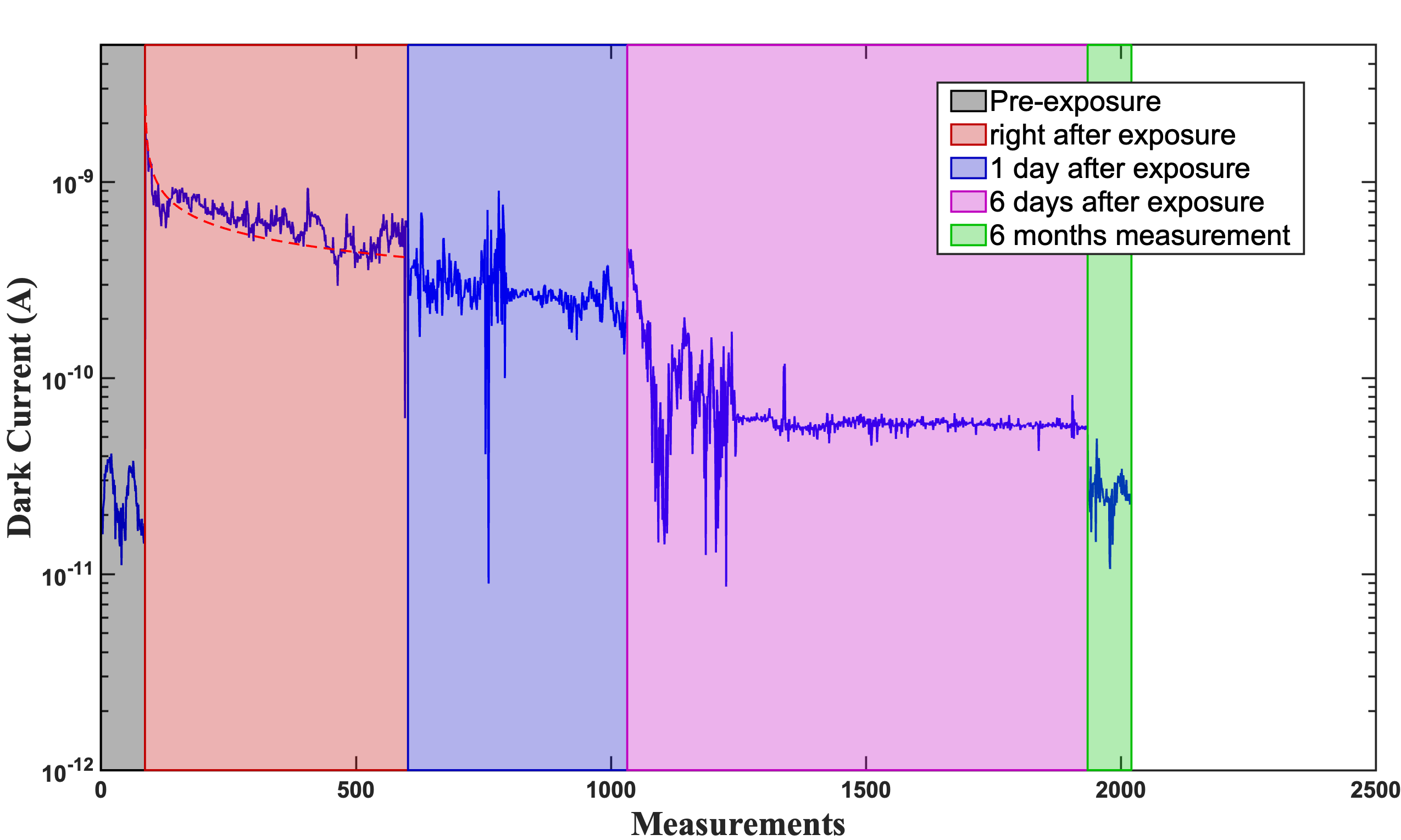}
    \caption{Quantum efficiency as a function of wavelength (left) and the evolution of dark current (right) before and after 23 hours
of lamp exposure.}
    \label{fig:lamp2}
\end{figure}

\begin{figure}[h]
    \centering
    \includegraphics[scale=.197]{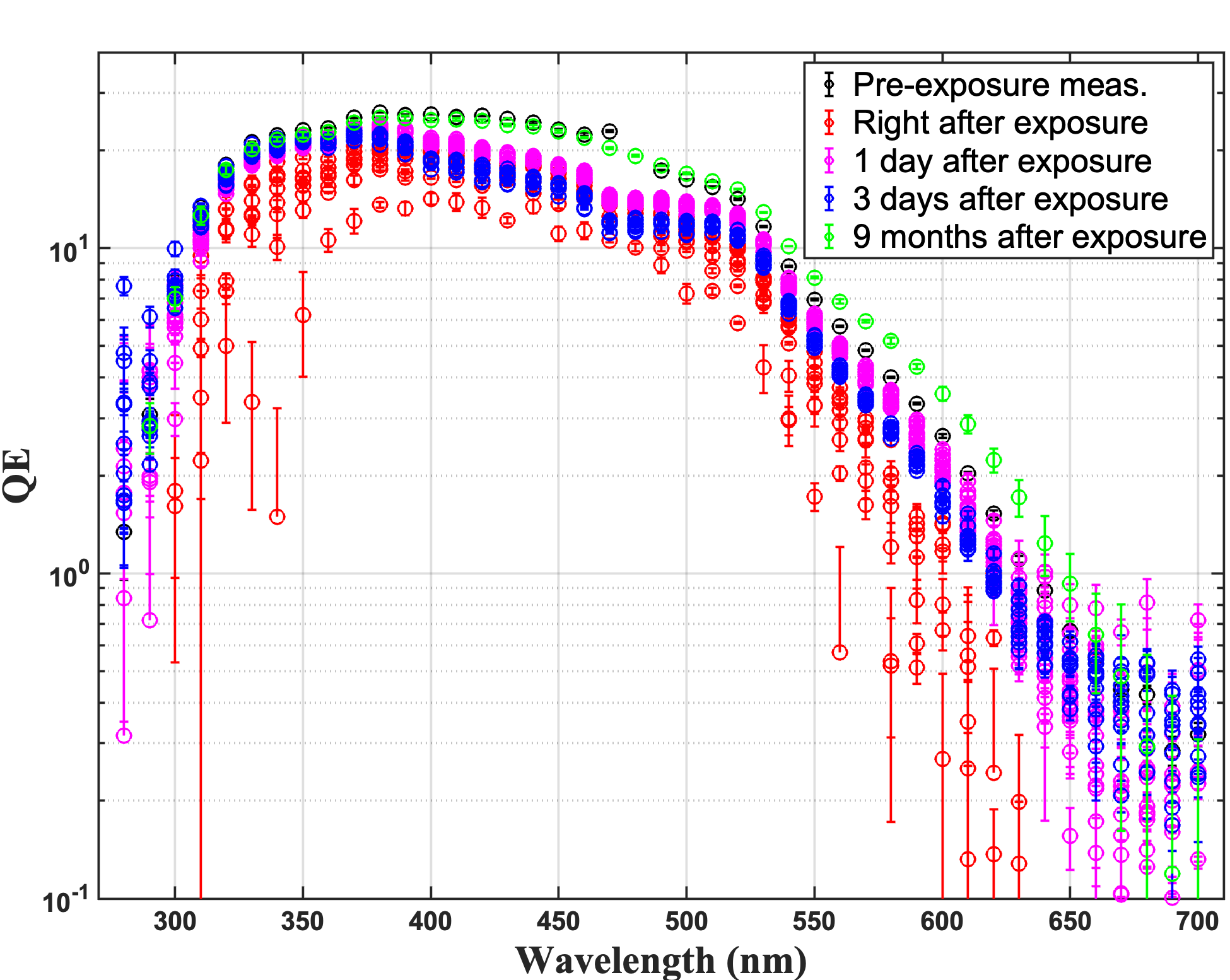}
    \includegraphics[scale=.197]{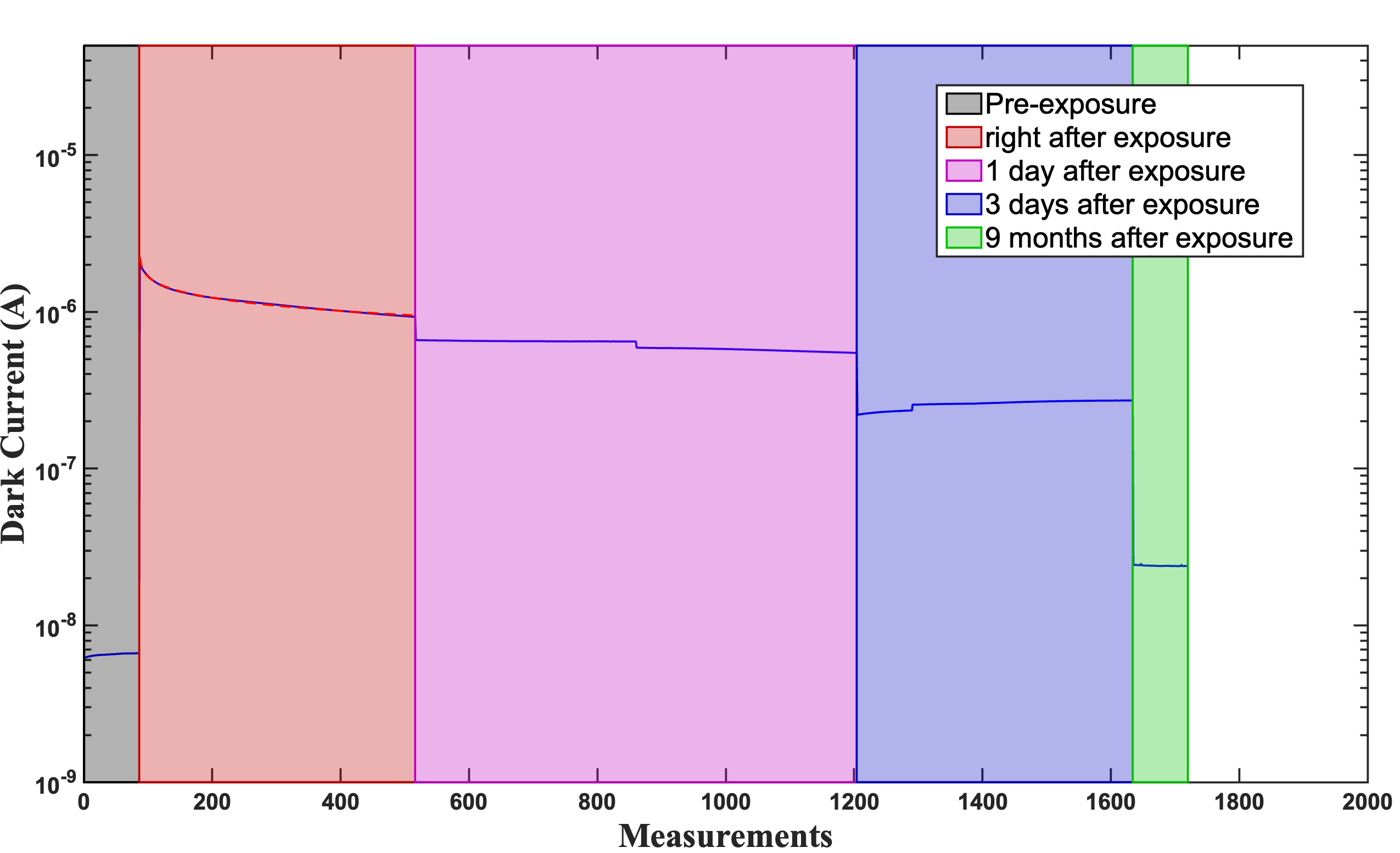}
    \caption{Quantum efficiency as a function of wavelength (left) and the evolution of dark current (right) before and after 70 hours
of lamp exposure.}
    \label{fig:lamp3}
\end{figure}

\figurename~\ref{fig:lamp2} and \figurename~\ref{fig:lamp3} present the QE and dark current trends for PMTs exposed for 23 and 70 hours, respectively. Both figures show a visible decrease in QE immediately after exposure, particularly in the wavelength range of 300 to 500 nm. This reduction in QE is accompanied by an increase in dark current, approximately two orders of magnitude higher than the pre-exposure levels, initially following a power-law decay.

Subsequent measurements, conducted over several days post-exposure (represented by differently coloured dots for QE and shaded areas for dark current), indicate a gradual recovery of QE toward pre-exposure levels, along with a corresponding reduction in dark current.

It is notable that, while QE values tend to return to their original levels relatively quickly, dark current takes longer to normalise. For example, in the case of the PMT exposed for 23 hours, the final measurements, taken six months post-exposure, show trends in QE and dark current consistent with the initial values (see \figurename~\ref{fig:lamp2}, right). As for the PMT exposed for 70 hours, it shows a decrease in dark current in the final measurement; however, the values remain approximately 0.7 orders of magnitude higher than those measured before the exposure (see \figurename~\ref{fig:lamp3}, right).



\subsection{Thermal exposure}
\label{th}
A sample of PMTs were subjected to thermal stress at various temperatures (ranging from 50$^{\circ}~$C to 210$^{\circ}~$C) and for different durations. Measurements of QE and dark current taken after exposure revealed four distinct scenarios, depending on the temperature range and exposure duration:

\begin{itemize}
\item \textbf{First scenario}: no QE degradation. For PMTs exposed at temperatures of 50$^{\circ}~$C and 70$^{\circ}~$C for up to 4 days, there was no noticeable degradation in QE, as illustrated in \figurename~\ref{fig:therm1}. This figure shows the QE and dark current distributions before and after exposing a PMT to 70$^{\circ}~$C for four days.

\begin{figure}[h]
    \centering
    \includegraphics[scale=.197]{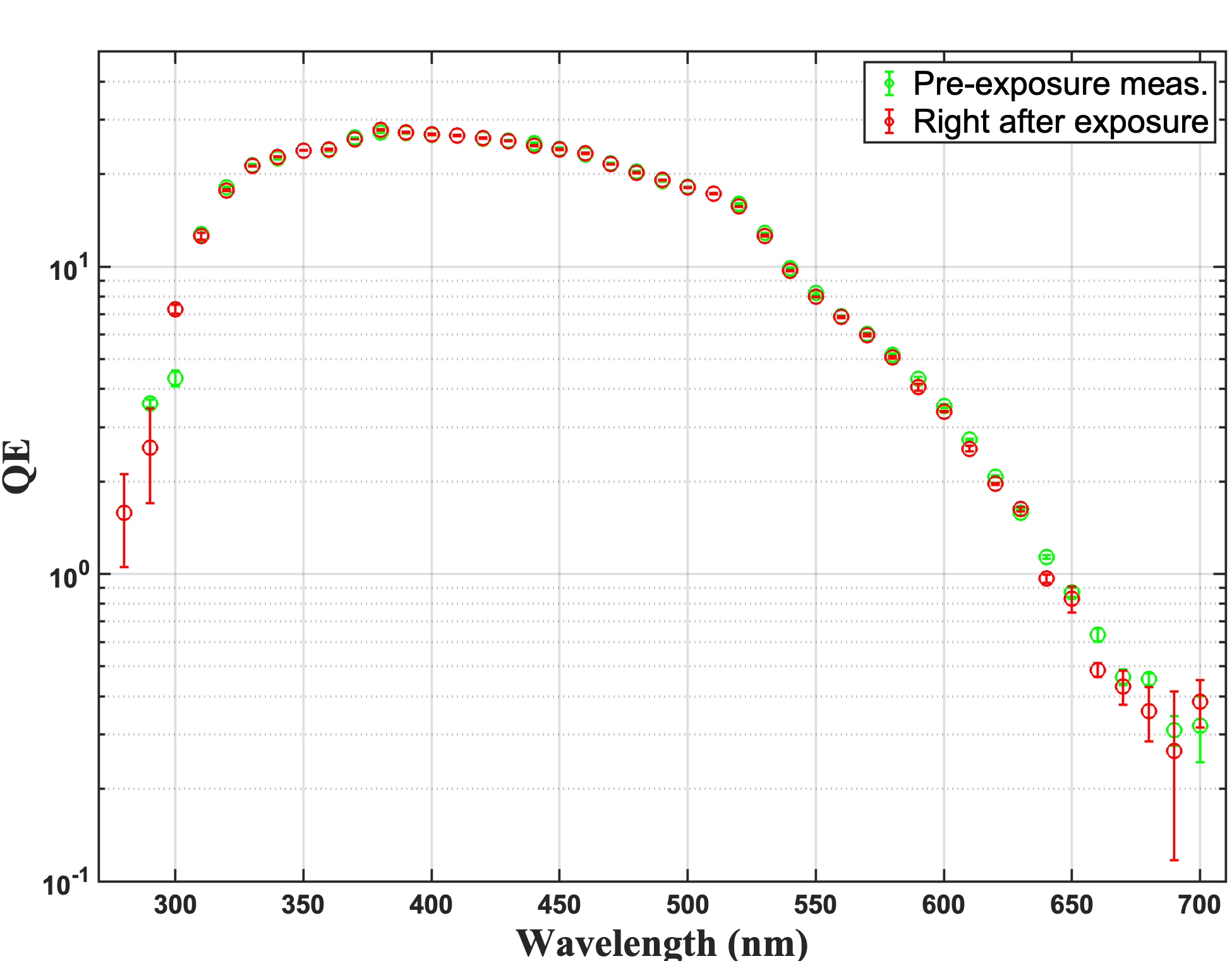}
    \includegraphics[scale=.197]{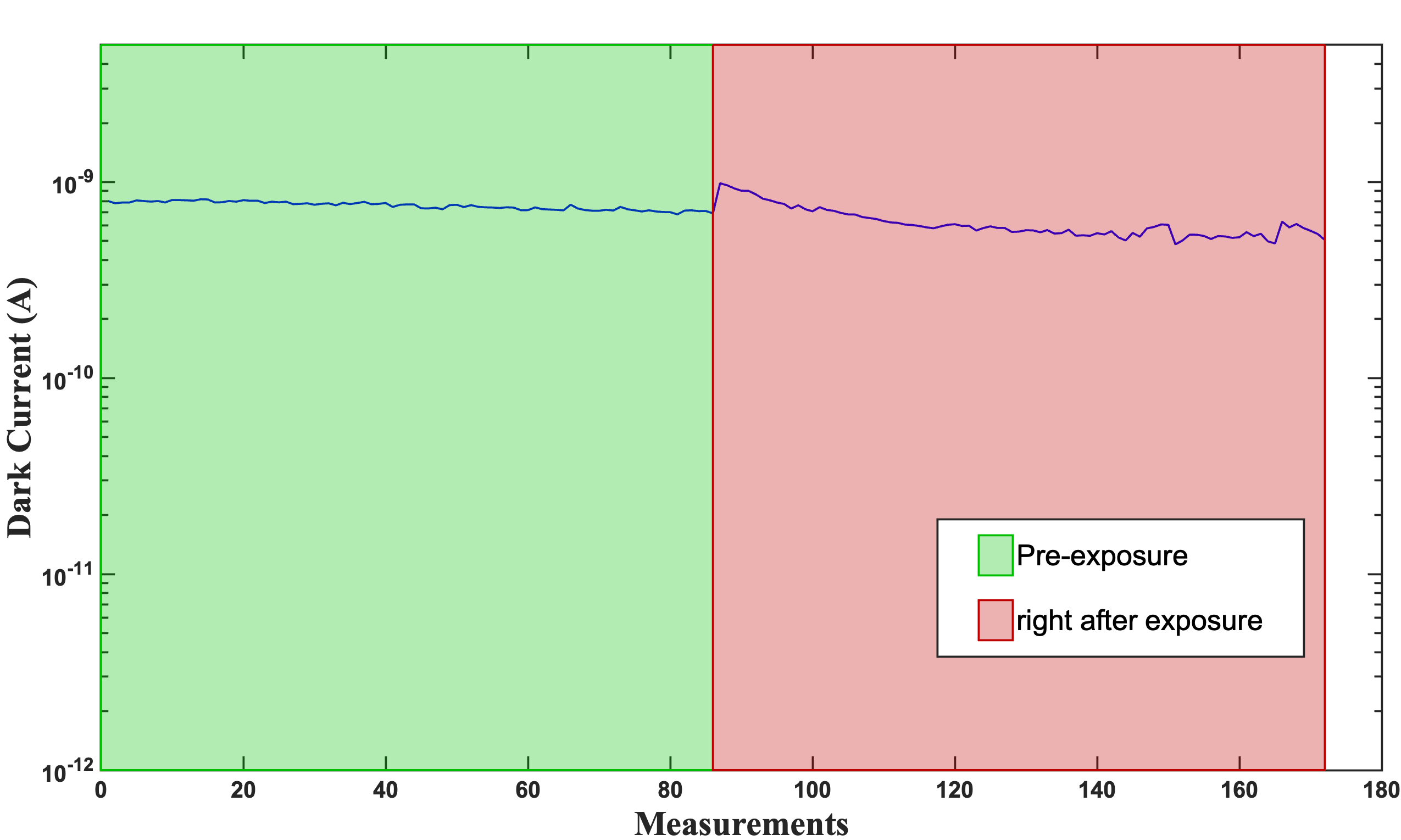}
    \caption{Quantum efficiency as a function of wavelength (left) and the evolution of dark current (right) before and after four days of thermal exposure at 70$^{\circ}$C.}
    \label{fig:therm1}
\end{figure}

\item \textbf{Second scenario}: reversible degradation of QE in the range between 350 and 550 nm. Similar to the effects observed in PMTs exposed to lamp light, reversible QE degradation was noted under the following conditions:

\begin{itemize}
\item[$\circ$] 90$^{\circ}~$C for two days and 180$^{\circ}~$C for one day (\figurename~\ref{fig:therm2});
\item[$\circ$] 180$^{\circ}~$C for four days (\figurename~\ref{fig:therm3});
\item[$\circ$] 210$^{\circ}~$C for seventeen hours (\figurename~\ref{fig:therm4}).
\end{itemize}

\begin{figure}[h]
    \centering
    \includegraphics[scale=.197]{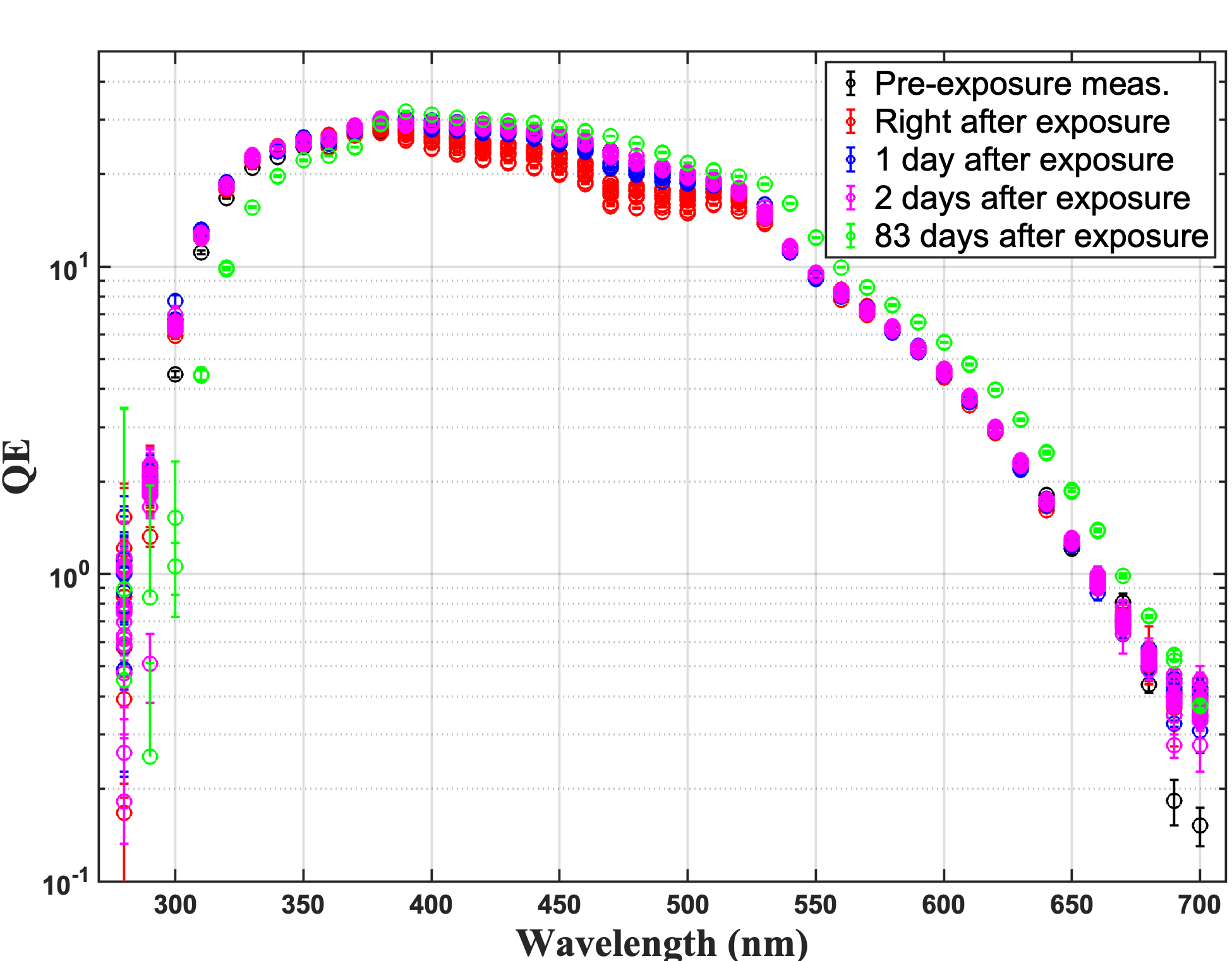}
    \includegraphics[scale=.197]{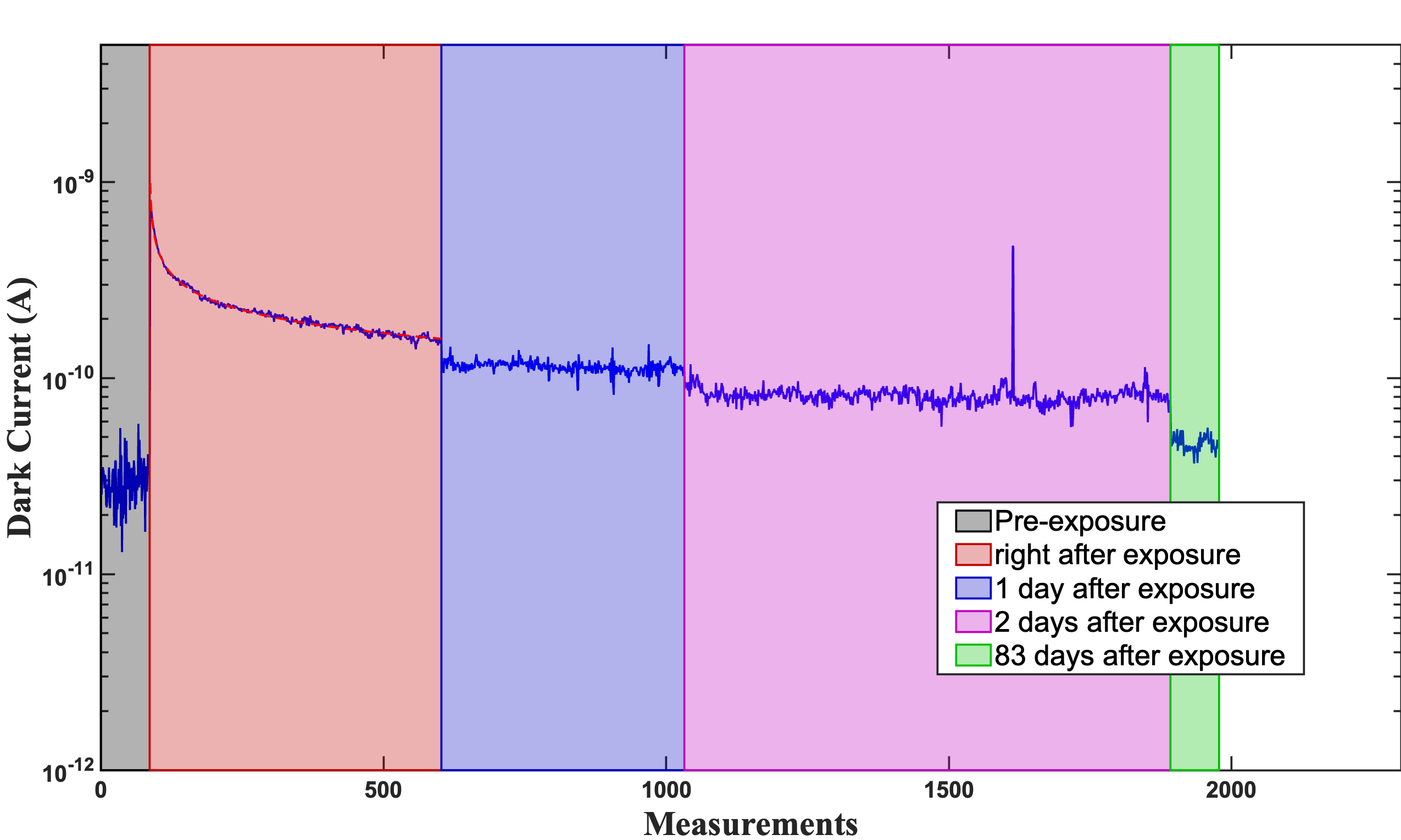}
    \caption{Quantum efficiency as a function of wavelength (left) and the evolution of dark current (right) before and after two days of thermal exposure at 90$^{\circ}$C and one day at 180$^{\circ}$C.}
    \label{fig:therm2}
\end{figure}

\begin{figure}[h]
    \centering
    \includegraphics[scale=.197]{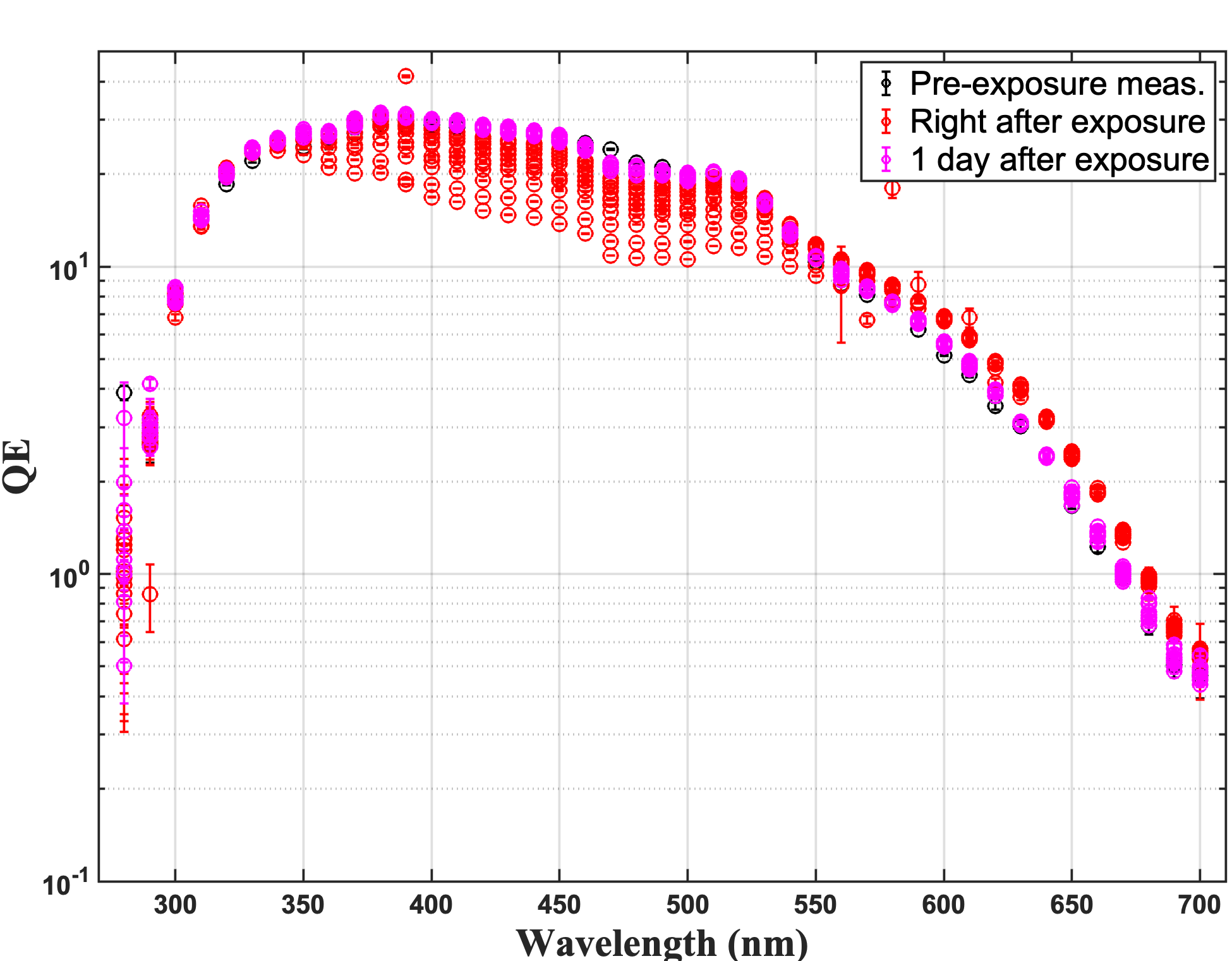}
    \includegraphics[scale=.197]{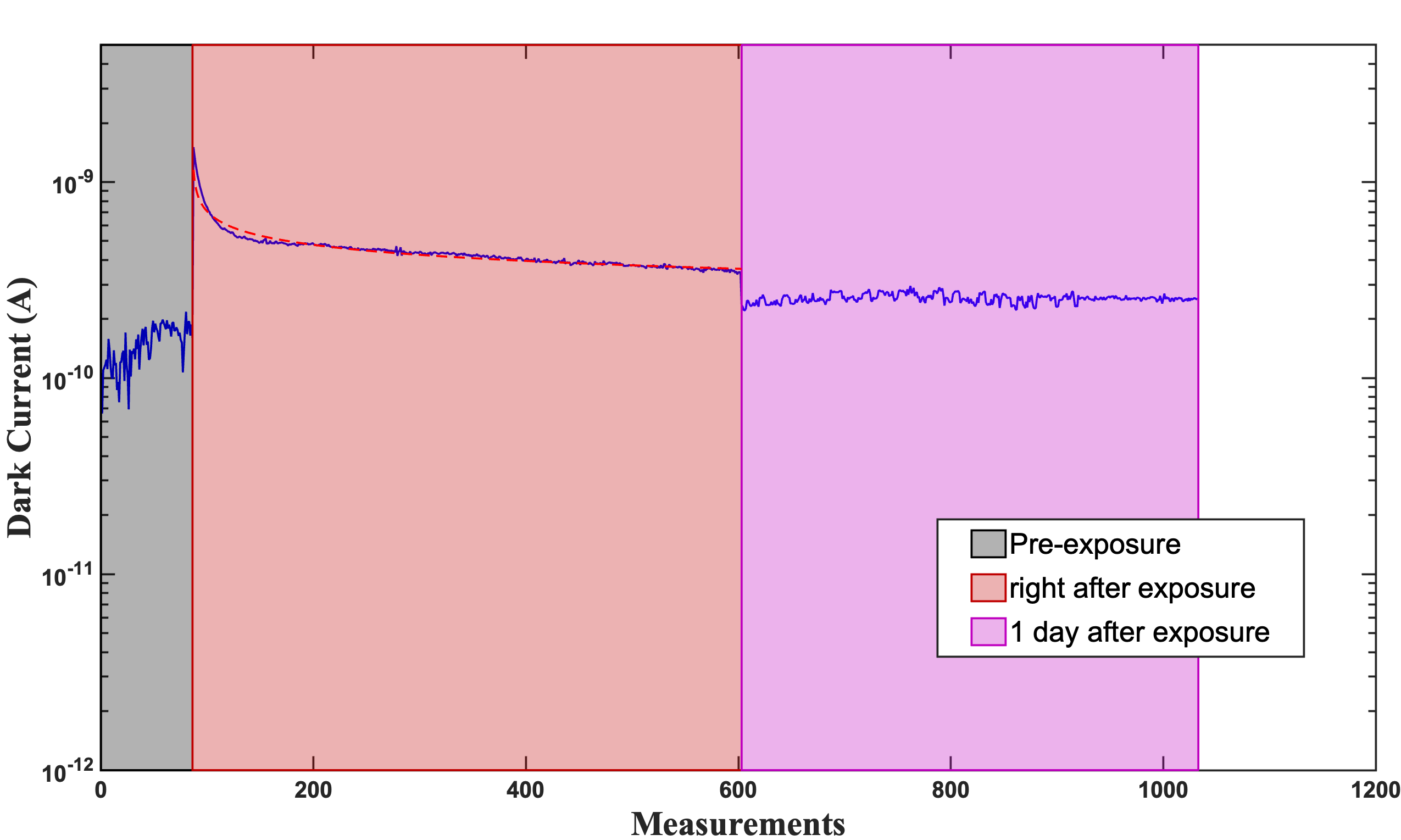}
    \caption{Quantum efficiency as a function of wavelength (left) and the evolution of dark current (right) before and after four days of thermal exposure at 180$^{\circ}$C.}
    \label{fig:therm3}
\end{figure}

\begin{figure}[h]
    \centering
    \includegraphics[scale=.197]{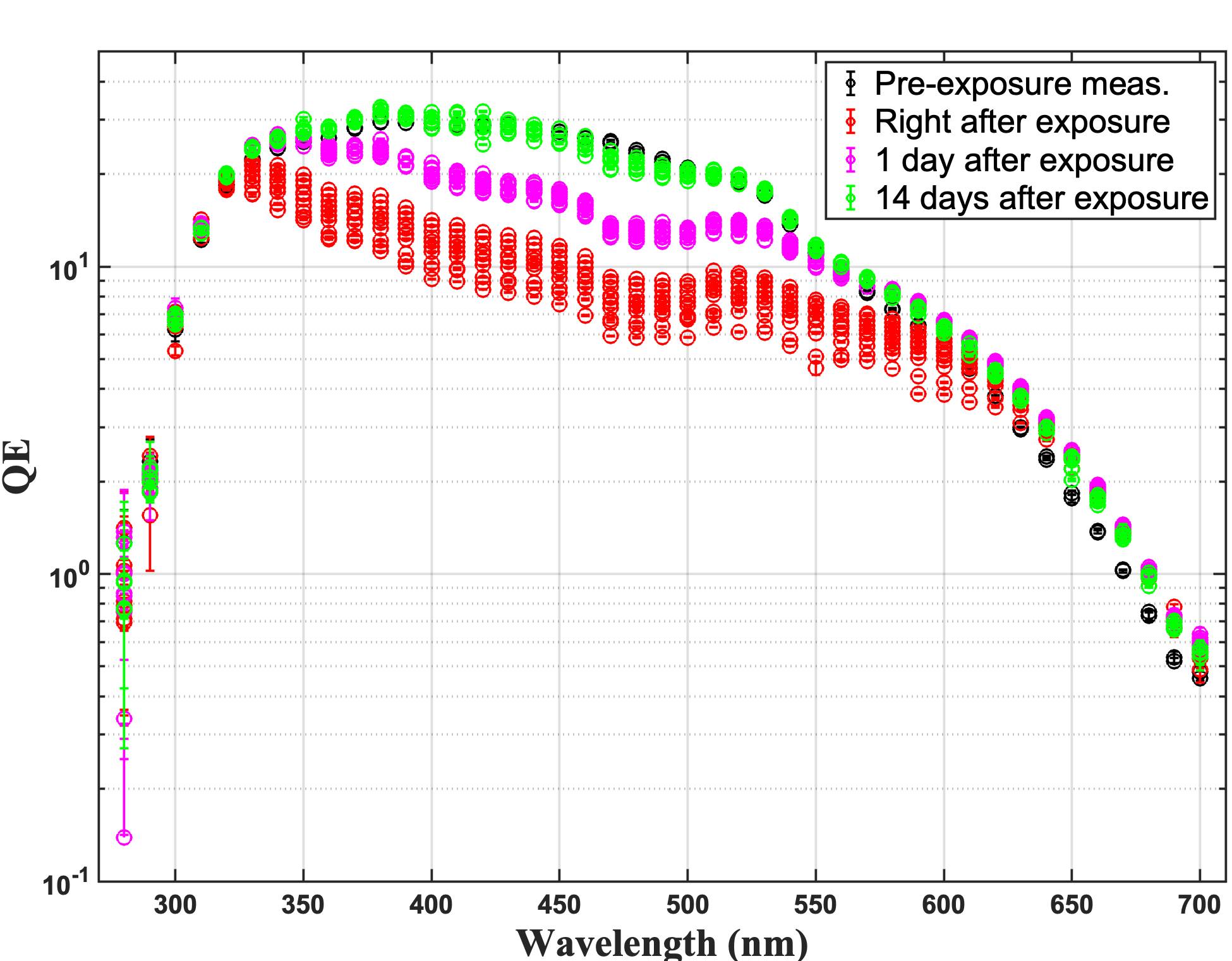}
    \includegraphics[scale=.197]{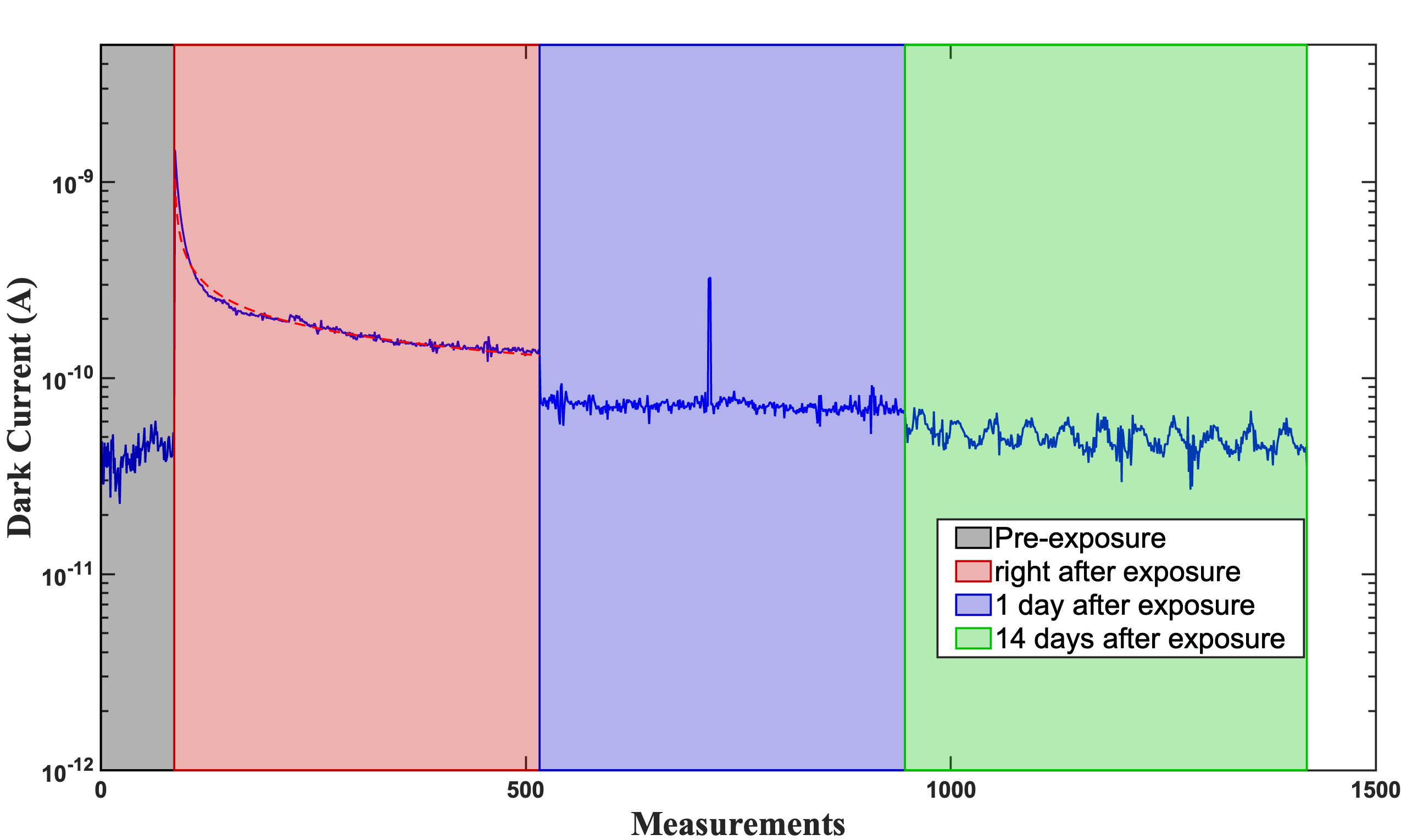}
    \caption{Quantum efficiency as a function of wavelength (left) and the evolution of dark current (right) before and after 17 hours of thermal exposure at 210$^{\circ}$C.}
    \label{fig:therm4}
\end{figure}

\item \textbf{Third scenario}: QE degradation across the entire wavelength range. This was observed in PMTs exposed to:

\begin{itemize}
\item[$\circ$] 190$^{\circ}~$C for four days (\figurename~\ref{fig:therm5});
\item[$\circ$] 210$^{\circ}~$C for 24 hours (\figurename~\ref{fig:therm6});
\item[$\circ$] 210$^{\circ}~$C for 28 hours (\figurename~\ref{fig:therm7}).
\end{itemize}

\begin{figure}[h]
    \centering
    \includegraphics[scale=.197]{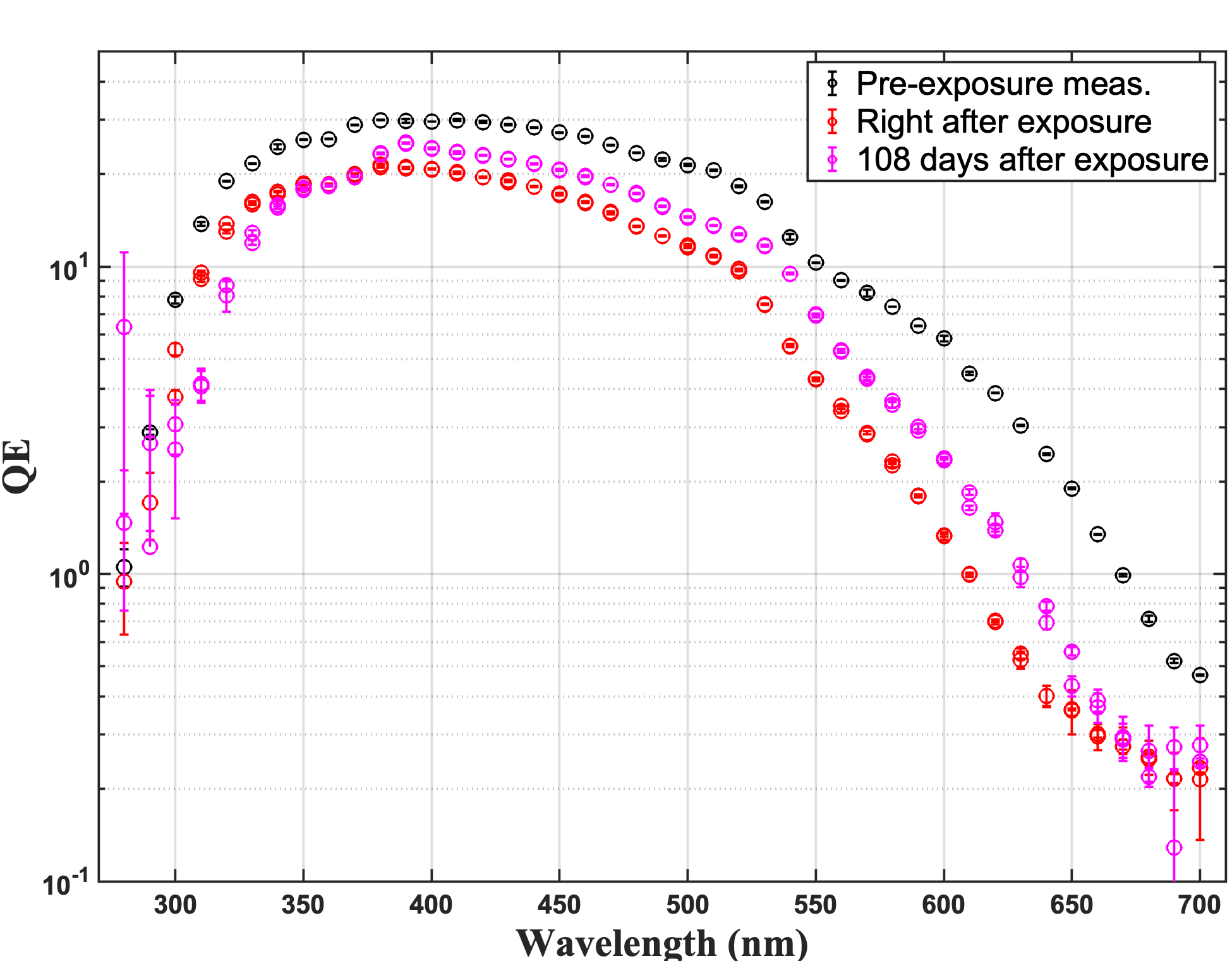}
    \includegraphics[scale=.197]{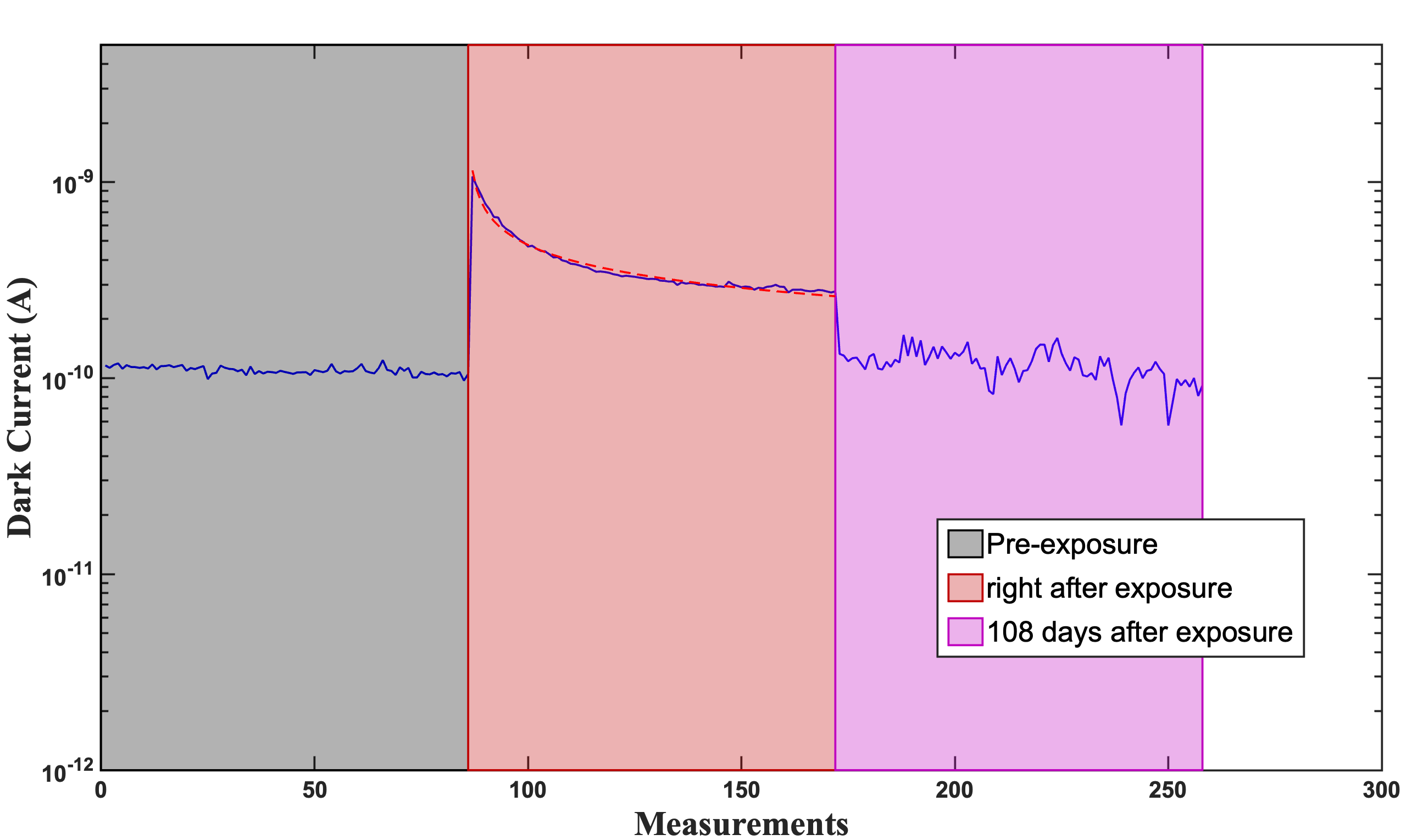}
    \caption{Quantum efficiency as a function of wavelength (left) and the evolution of dark current (right) before and after four days of thermal exposure at 190$^{\circ}$C.}
    \label{fig:therm5}
\end{figure}

\begin{figure}[h]
    \centering
    \includegraphics[scale=.197]{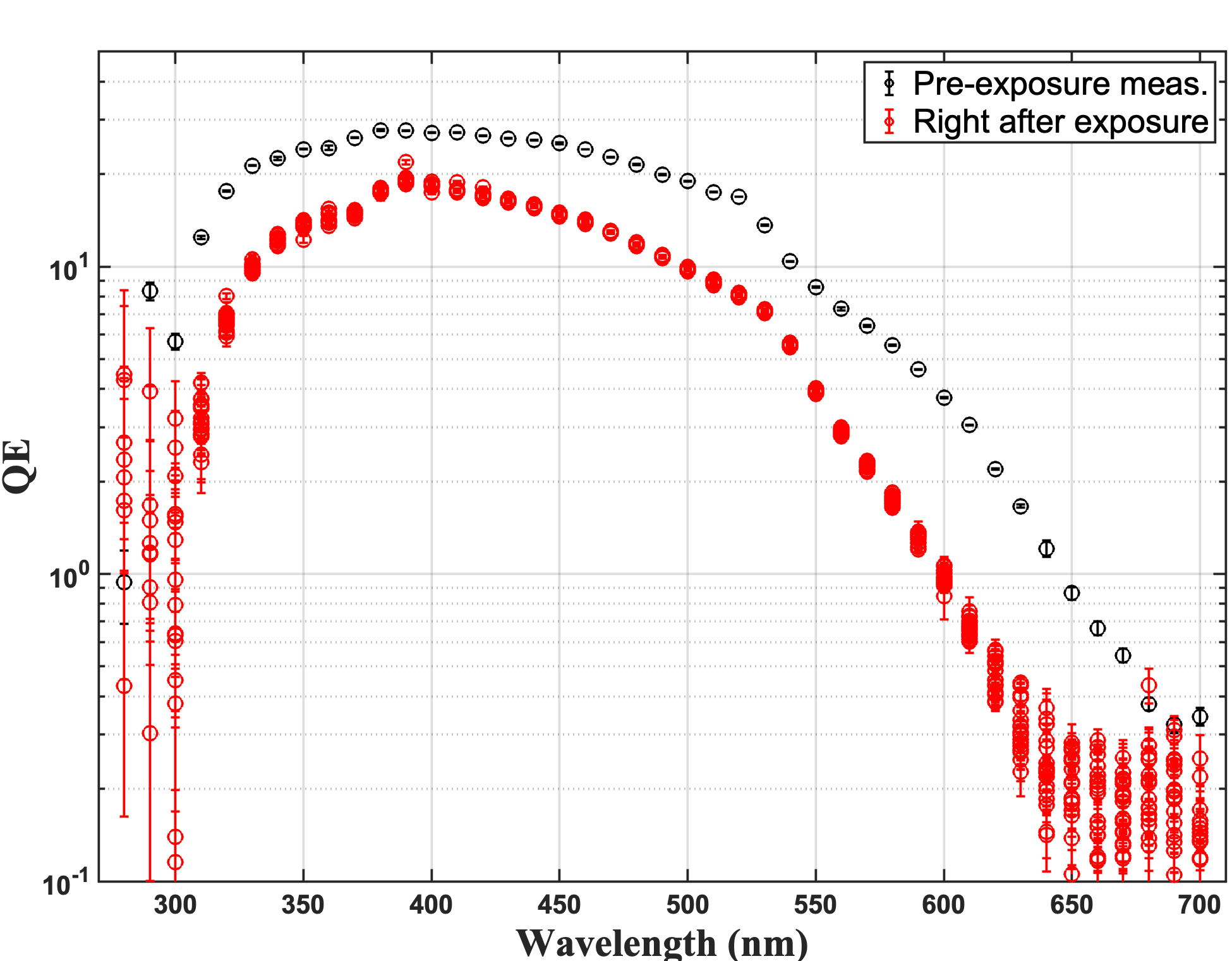}
    \includegraphics[scale=.197]{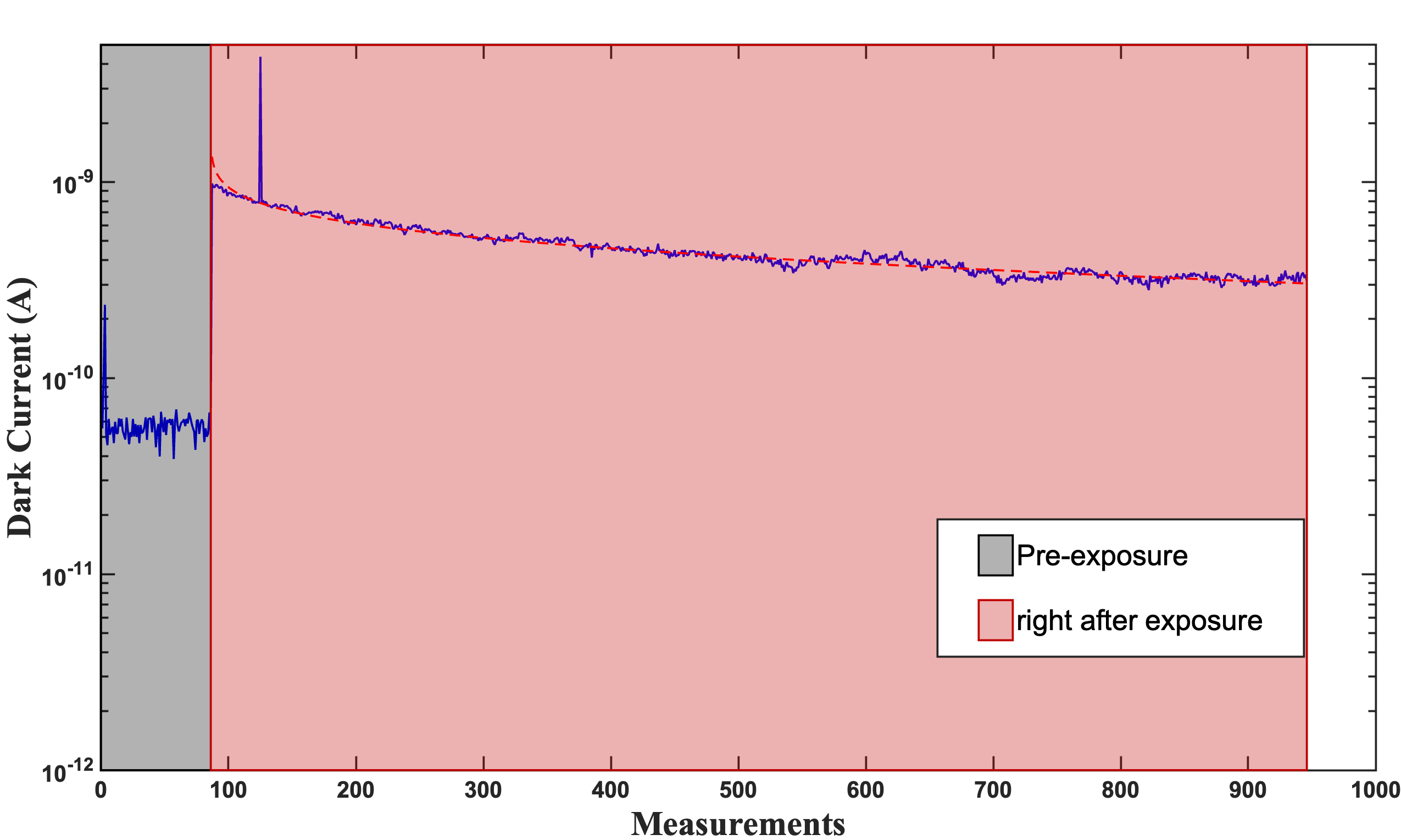}
    \caption{Quantum efficiency as a function of wavelength (left) and the evolution of dark current (right) before and after one day of thermal exposure at 210$^{\circ}$C.}
    \label{fig:therm6}
\end{figure}

\begin{figure}[h]
    \centering
    \includegraphics[scale=.197]{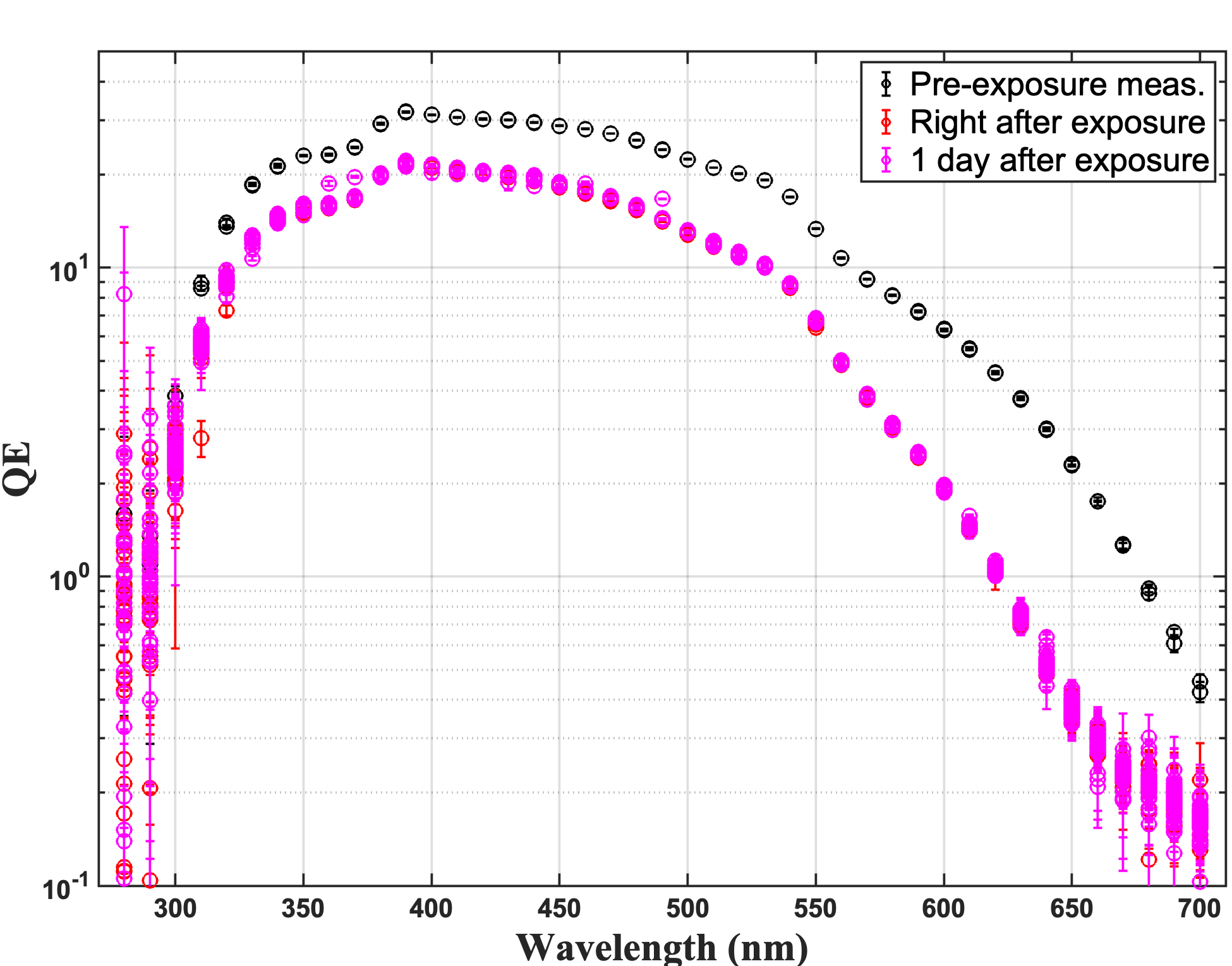}
    \includegraphics[scale=.197]{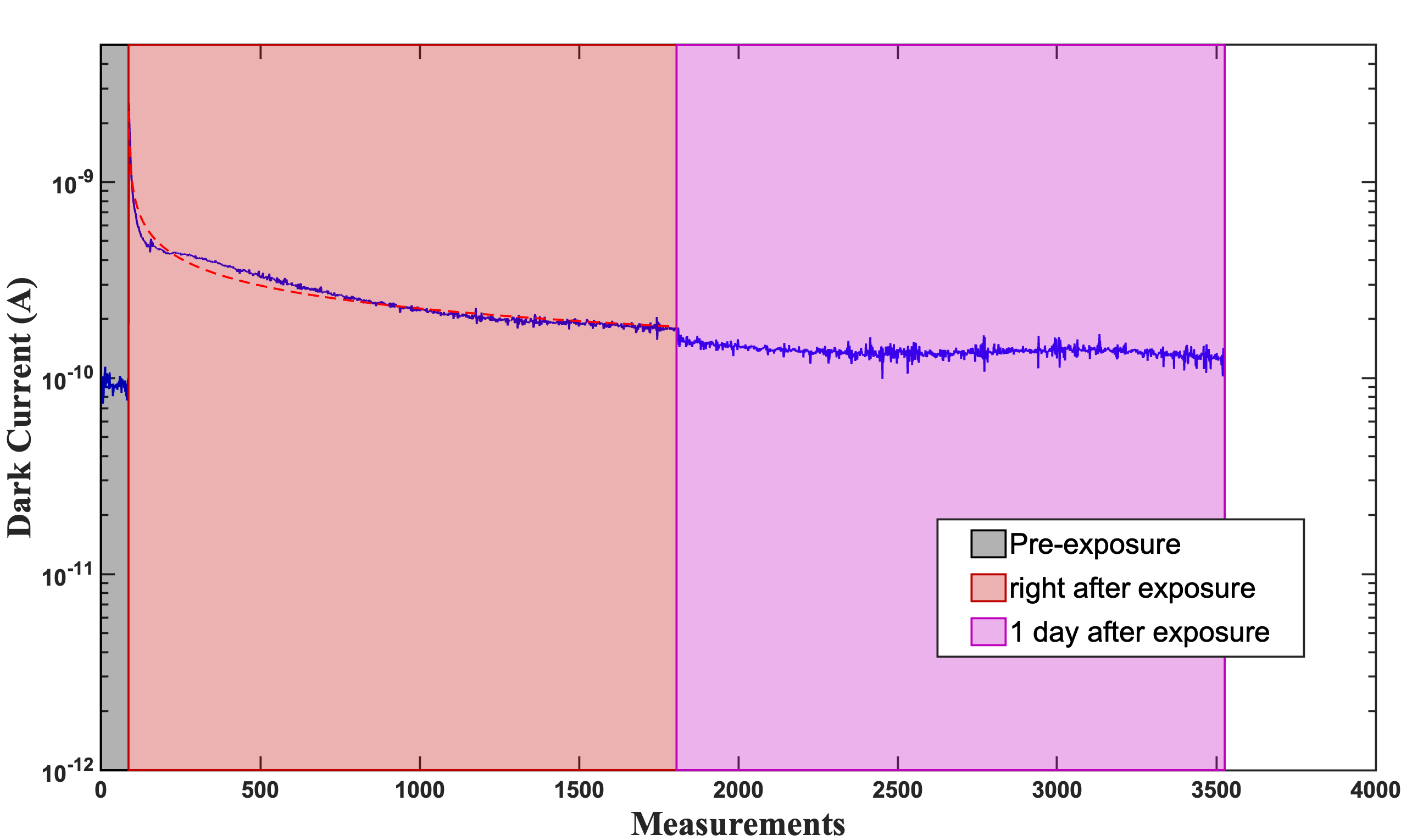}
    \caption{Quantum efficiency as a function of wavelength (left) and the evolution of dark current (right) before and after 28 hours of thermal exposure at 210$^{\circ}$C.}
    \label{fig:therm7}
\end{figure}


In these cases, the reduction in performance appears to be either permanent or characterized by a very slow recovery to pre-exposure conditions, as can be observed, for example, in \figurename~\ref{fig:therm5}, where the QE measurement taken 108 days after exposure shows only a slight improvement compared to the measurements taken immediately after exposure.

\item \textbf{Fourth scenario}: permanent damage. This scenario was observed only once. \figurename~\ref{fig:therm8} shows the QE and dark current measurements of a PMT exposed for five days at 180$^{\circ}~$C  and for one day at 210$^{\circ}~$C. In this case, the QE distribution is completely suppressed, and measurements taken two months after the exposure show no signs of recovery, while the dark current exhibits a continuous decrease.

\begin{figure}[h]
    \centering
    \includegraphics[scale=.197]{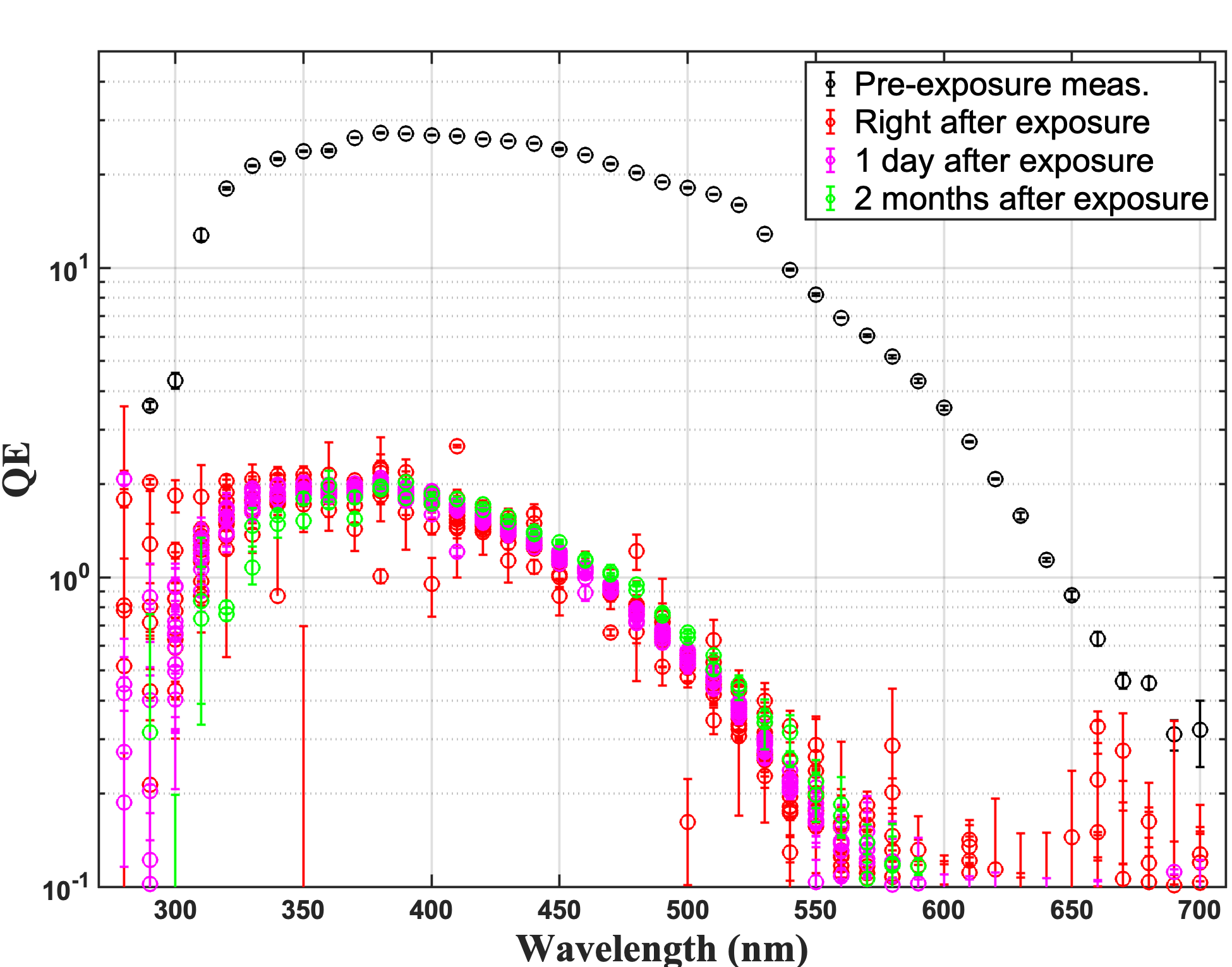}
    \includegraphics[scale=.197]{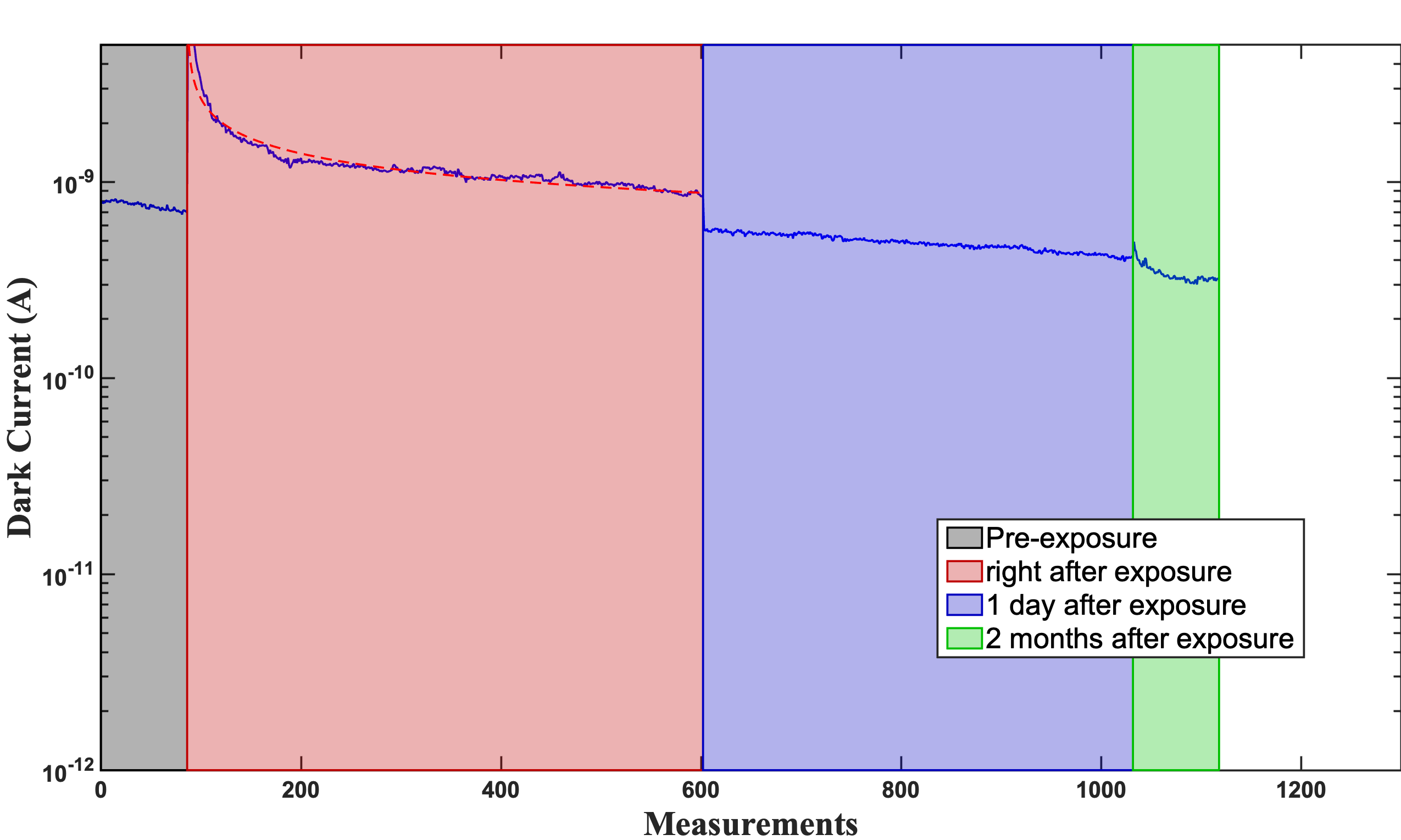}
    \caption{Quantum efficiency as a function of wavelength (left) and the evolution of dark current (right) before and after five days of thermal exposure at 180$^{\circ}$C and one day at 210$^{\circ}$C.}
    \label{fig:therm8}
\end{figure}

\end{itemize}


In the first three scenarios, the dark current increased by more than an order of magnitude compared to pre-exposure values, displaying a power-law decay in the initial stages. Additionally, the dark current takes much longer to return to pre-exposure values compared to QE. For example, in the PMT exposed to a temperature of 90$^{\circ}~$C for two days and then to 180$^{\circ}~$C on the following day (\figurename~\ref{fig:therm2}), the final set of measurements, taken 83 days after exposure, shows dark current values approximately 67$\%$ higher than pre-exposure levels. In contrast, in the fourth scenario, a continuous decrease is observed, with the usual power-law decay in measurements taken immediately after exposure.

\begin{figure}[h]
    \centering
        \begin{minipage}{.42\textwidth}
        \centering
        \includegraphics[width=\textwidth]{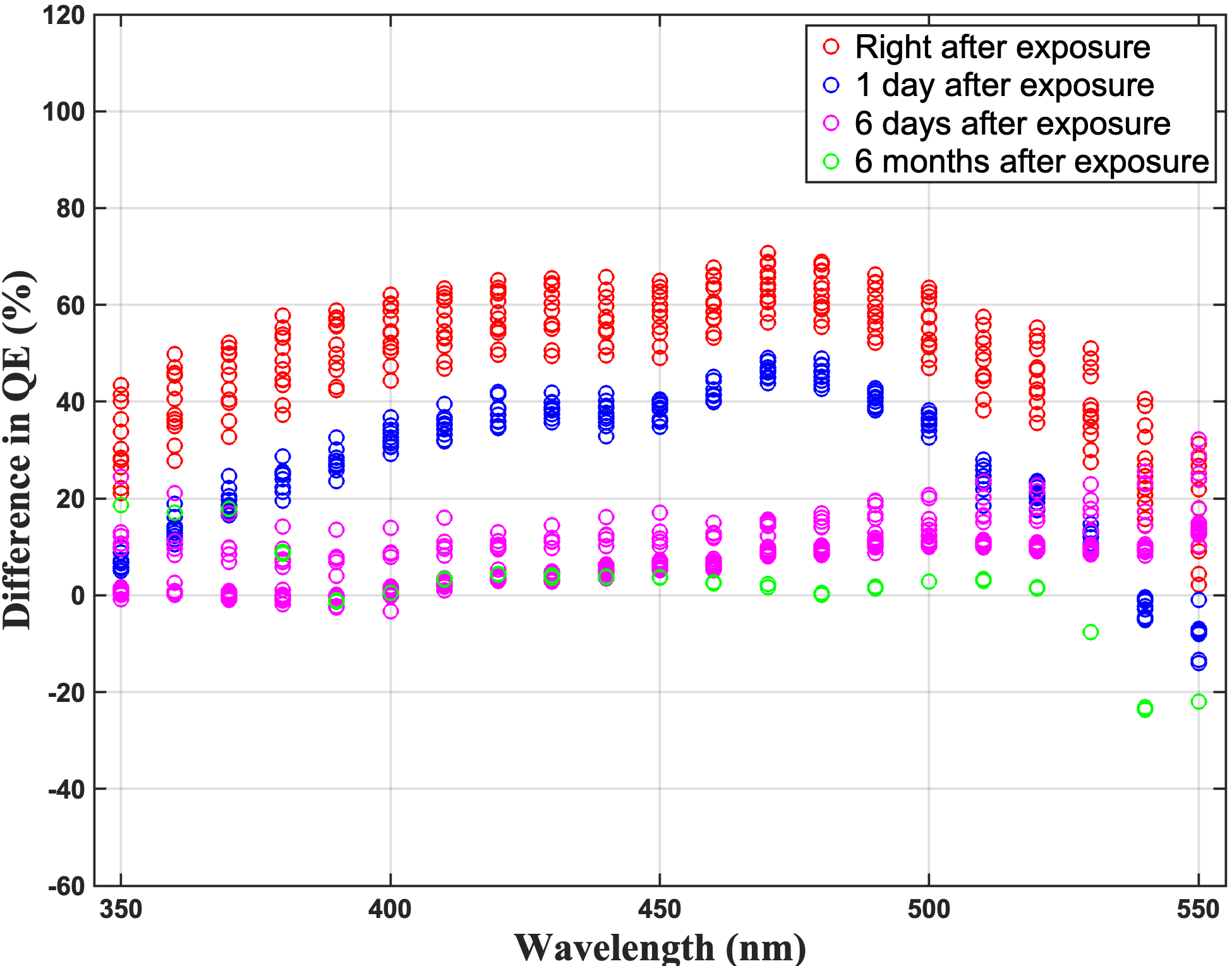}
        (a)
    \end{minipage}%
    \hspace{1cm}
    \begin{minipage}{.42\textwidth}
        \centering
        \includegraphics[width=\textwidth]{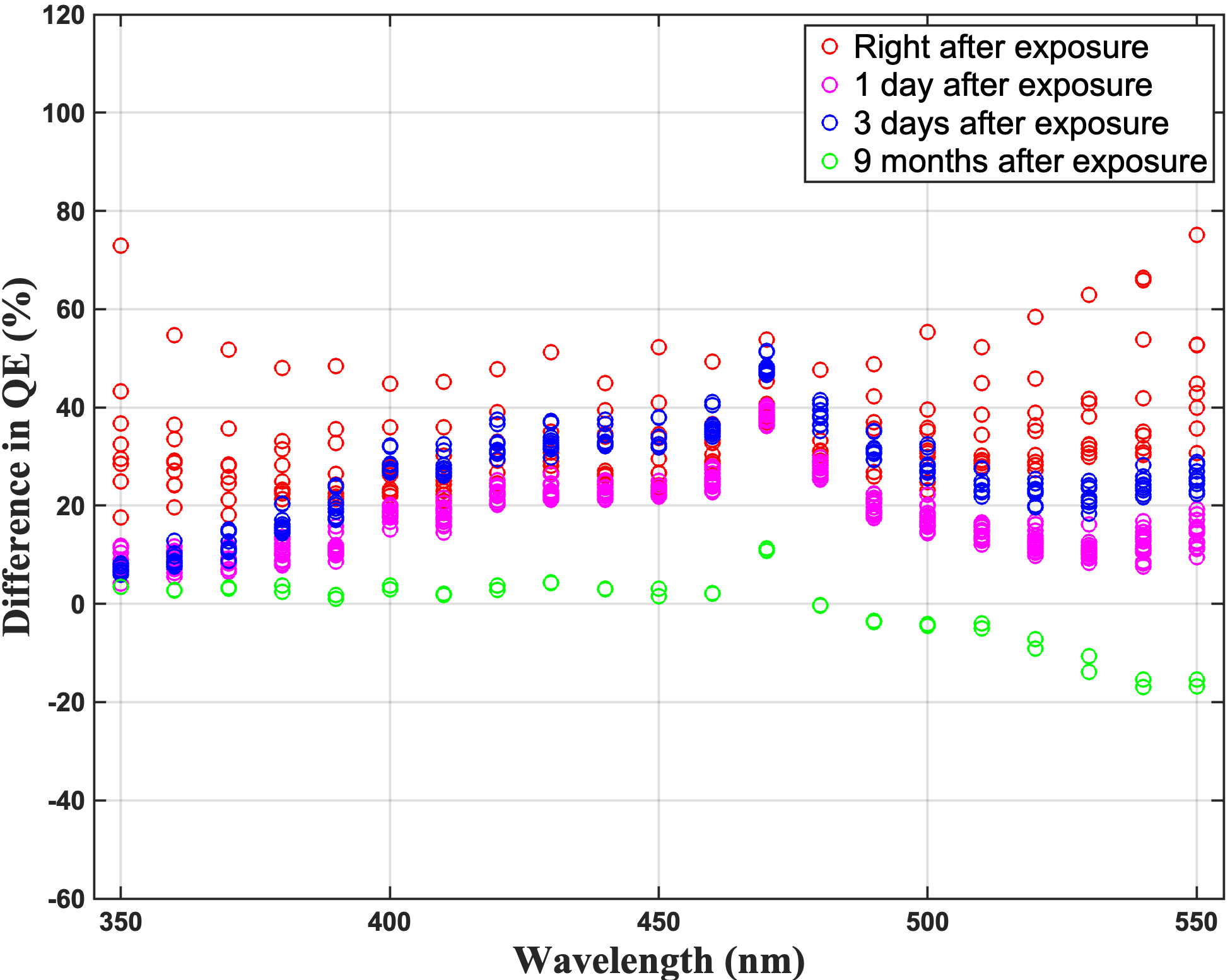}
        (b)
    \end{minipage}%

    \caption{Evolution of the percentage change in QE as a function of wavelength for PMTs exposed to lamp light for 23 hours (a) and 70 hours (b), respectively}
    \label{fig:diff1}
\end{figure}

\begin{figure}[h]
    \centering
    \begin{minipage}{.42\textwidth}
        \centering
        \includegraphics[width=\textwidth]{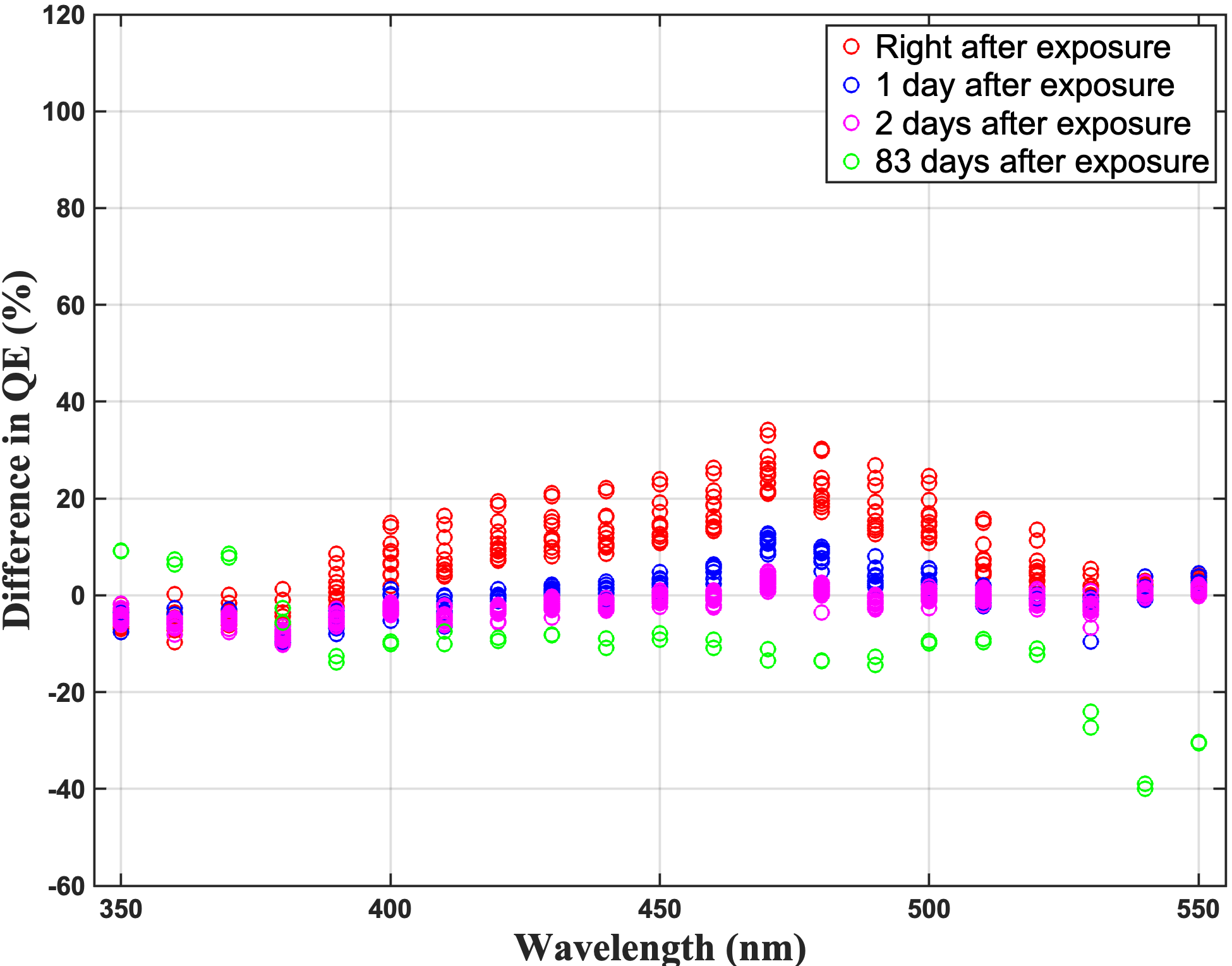}
        (a)
    \end{minipage}%
    \hspace{1cm}
    \begin{minipage}{.42\textwidth}
        \centering
        \includegraphics[width=\textwidth]{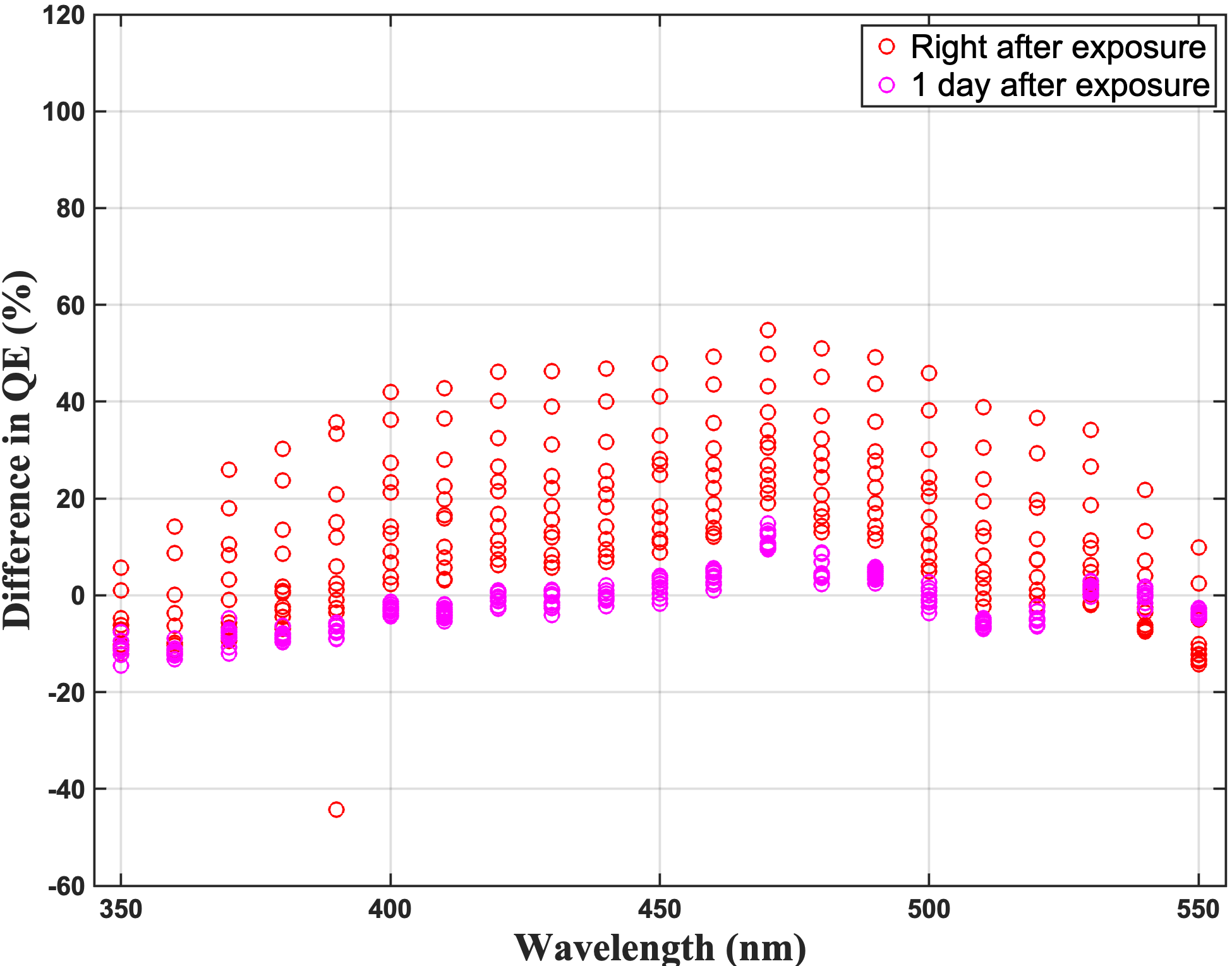}
        (b)
    \end{minipage}%

    \vspace{10pt} 

    \begin{minipage}{.42\textwidth}
        \centering
        \includegraphics[width=\textwidth]{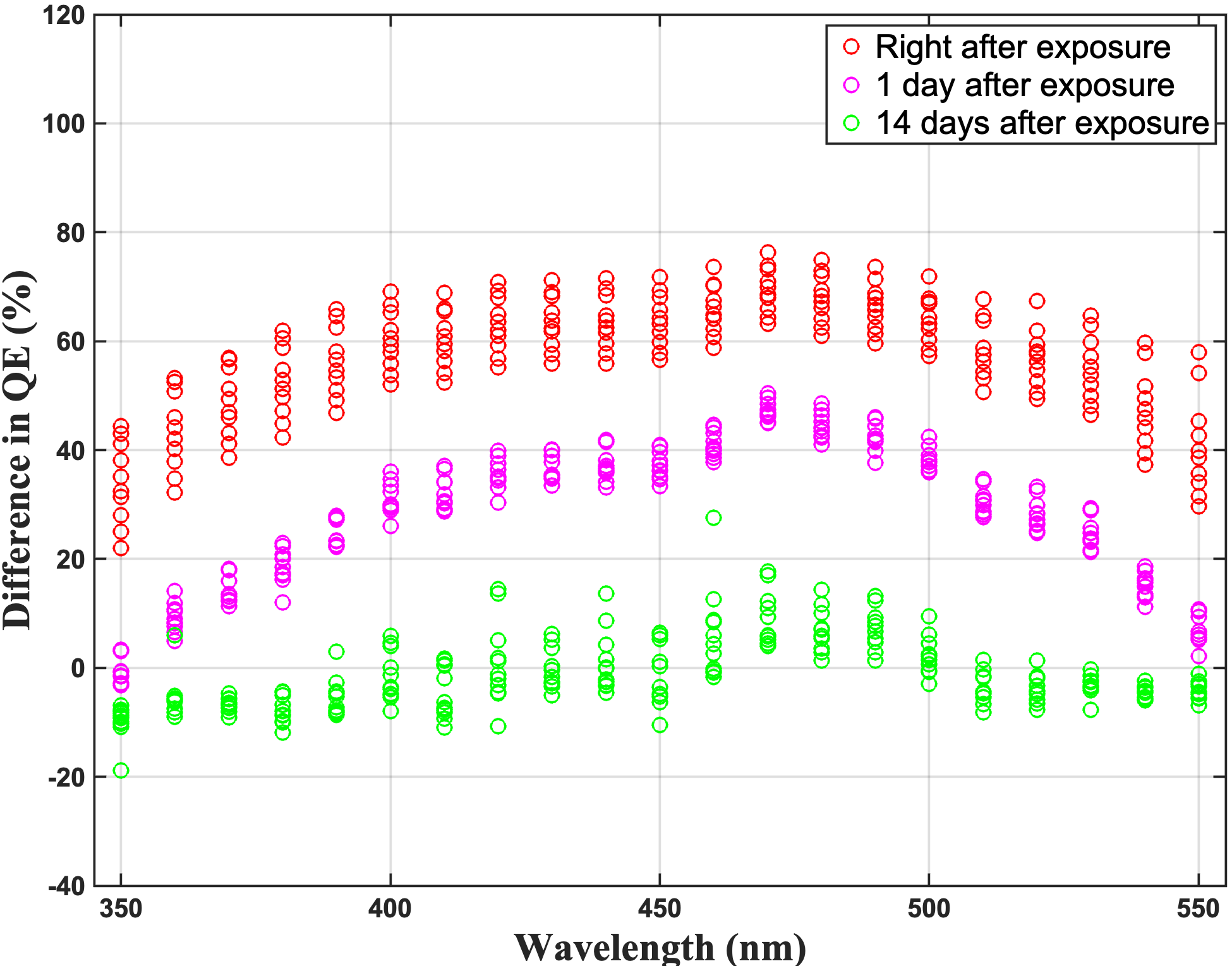}
        (c)
    \end{minipage}%
  
    \caption{Evolution of the percentage change in QE as a function of wavelength for PMTs exposed to thermal stress for two days at 90$^\circ$ and one at 180$^\circ$ (a); four days at 180$^\circ$ (b); 17 hours at 210$^\circ$ (c) respectively.}
    \label{fig:diff2}
\end{figure}

\section{Discussion}

All PMTs exposed to light and thermal stress for a sufficiently long duration showed a degradation in QE. In the case of light stress and thermal stress under the second scenario (as described in the previous section), this degradation was found to be reversible and mainly affected the wavelength range between 350 and 550 nm. This wavelength range is considered in  \figurename~\ref{fig:diff1} and \figurename~\ref{fig:diff2}, which illustrate the percentage change in QE before and after exposure for two PMTs subjected to 23 and 70 hours of light exposure from the lamp, as well as for the PMTs exposed to thermal stress under the following conditions: two days at 90°C and one day at 180°C; four days at 180°C; and 17 hours at 210°C. Depending on the case, a maximum variation between 60\% and 80\% can be observed, generally occurring around 470 nm.

\begin{figure}[h]
    \centering
    \begin{minipage}{0.42\textwidth}
        \centering
        \includegraphics[width=\textwidth]{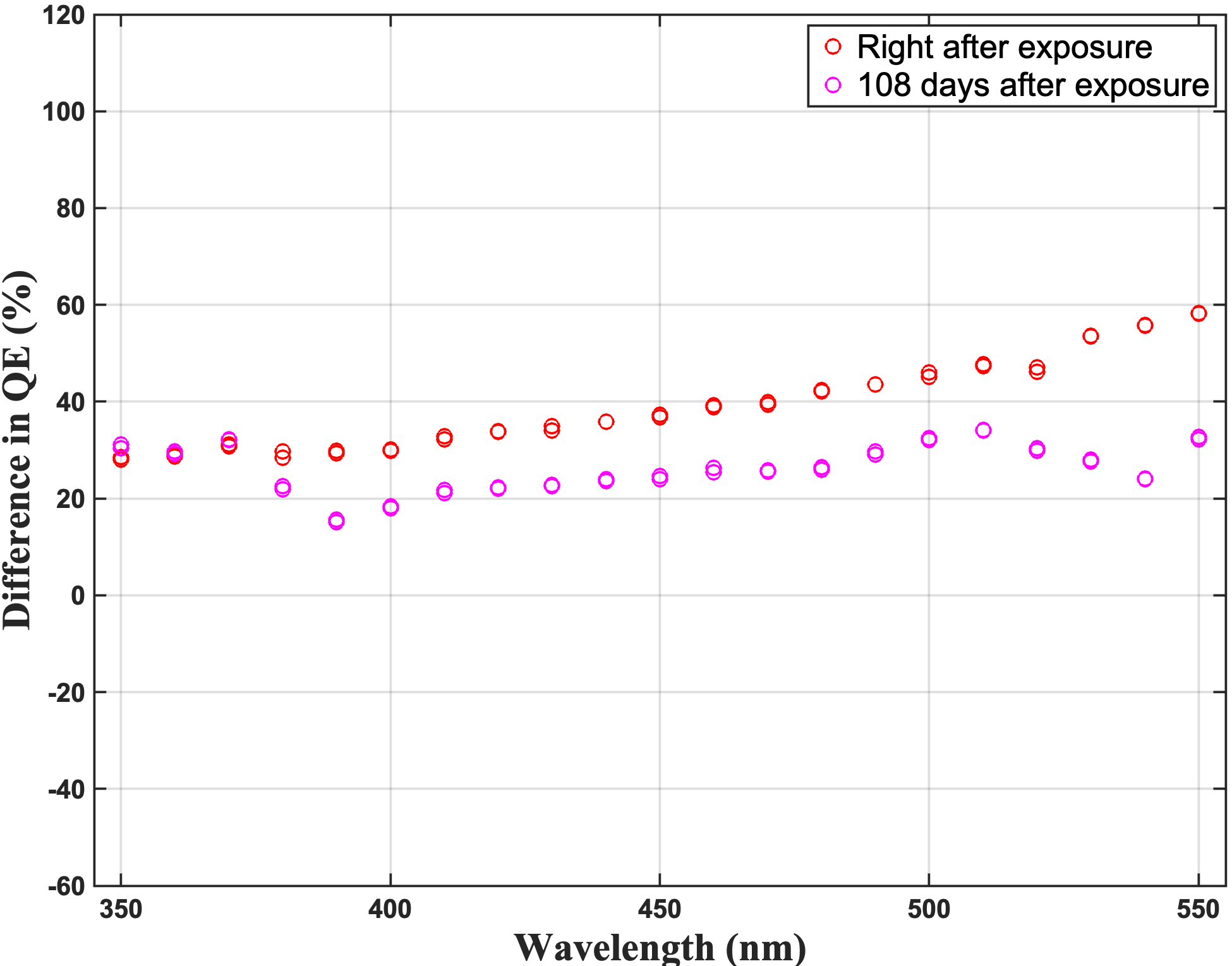}
        \\(a)
    \end{minipage}%
    \hspace{0.04\textwidth} 
    \begin{minipage}{0.42\textwidth}
        \centering
        \includegraphics[width=\textwidth]{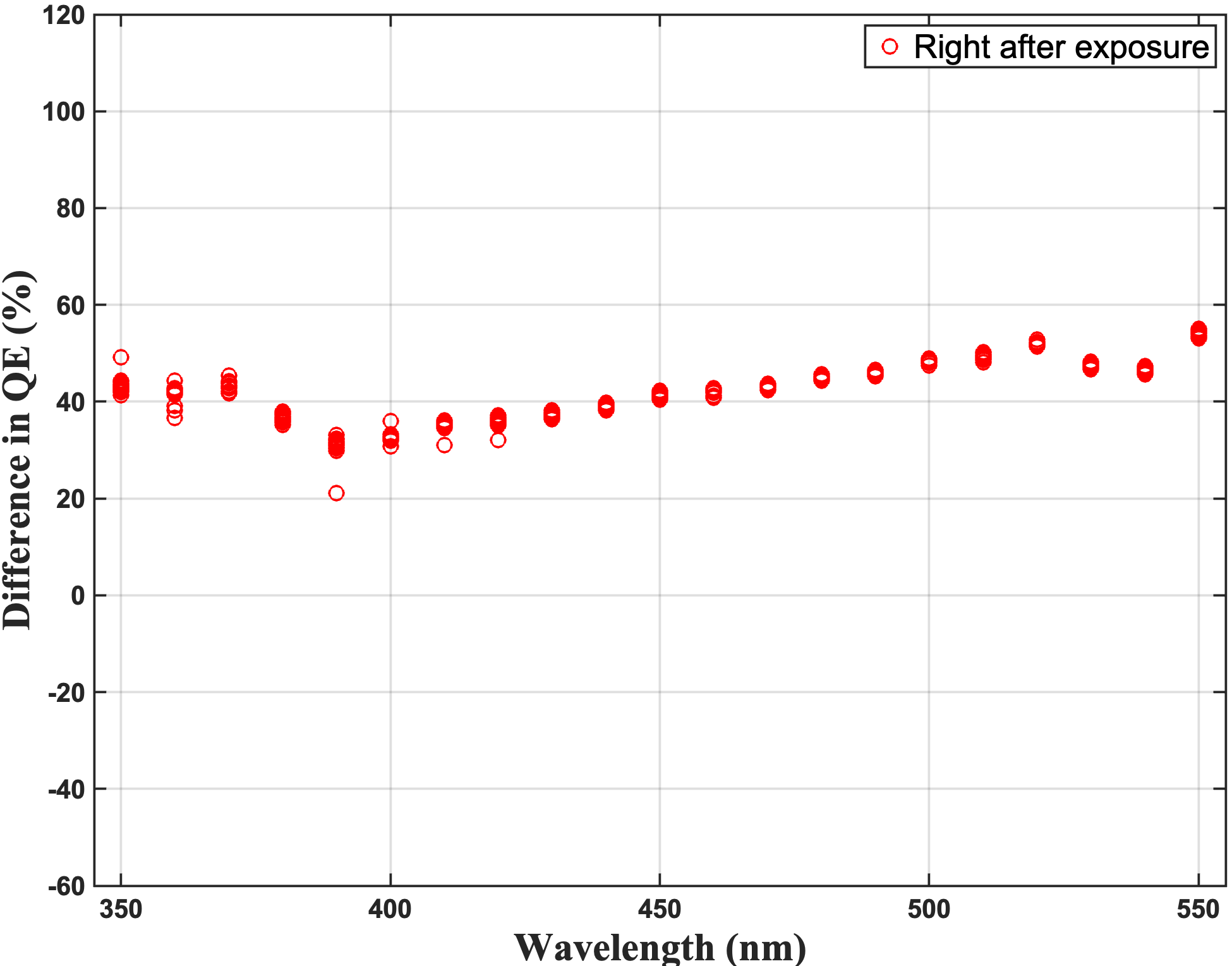}
        \\(b)
    \end{minipage}
    
    \vspace{10pt} 
    
    \begin{minipage}{0.42\textwidth}
        \centering
        \includegraphics[width=\textwidth]{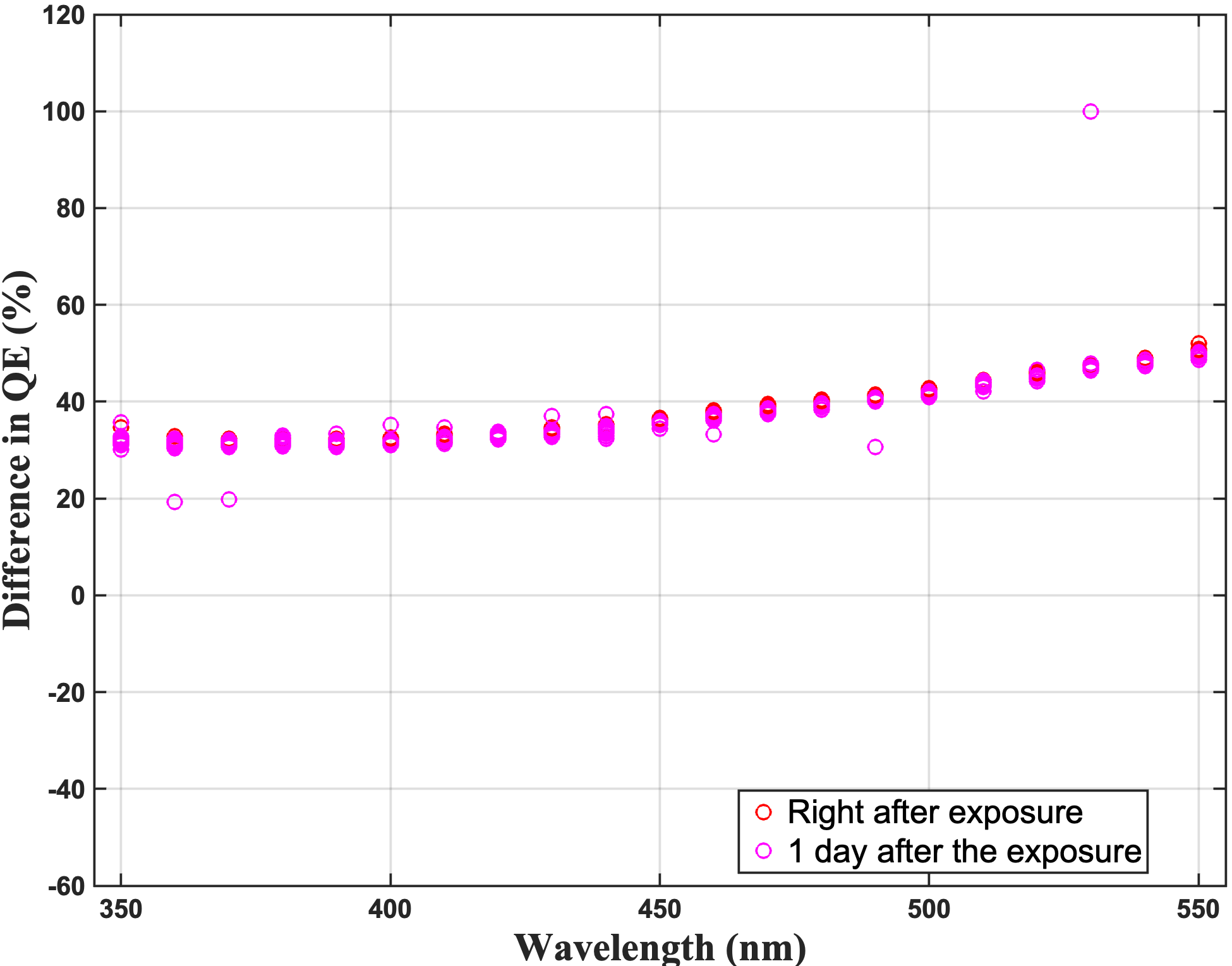}
        \\(c)
    \end{minipage}%
    \hspace{0.04\textwidth} 
    \begin{minipage}{0.42\textwidth}
        \centering
        \includegraphics[width=\textwidth]{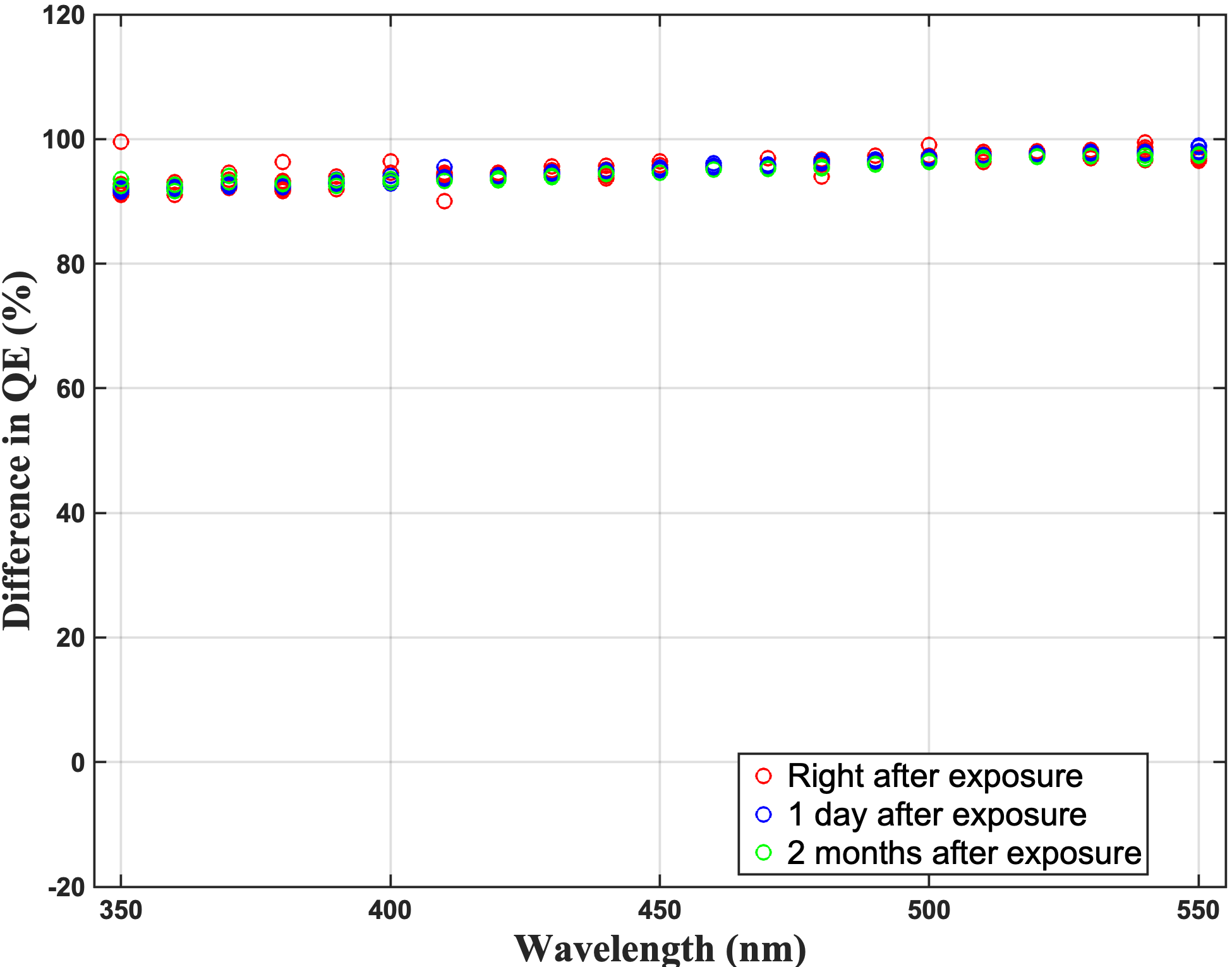}
        \\(d)
    \end{minipage}

    \caption{Evolution of the percentage change in QE as a function of wavelength for PMTs exposed to thermal stress for four days at 190$^\circ$ (a); one day at 210$^\circ$ (b); 28 hours at 210$^\circ$ (c); five days at 180$^\circ$ and one at 210$^\circ$ (d), respectively.}
    \label{fig:diff3}
\end{figure}

In \figurename~\ref{fig:diff3}, the percentage change in QE before and after exposure for PMTs exhibiting the behavior described in scenarios three and four of Section \ref{th} are shown. In all these cases, the variation is almost constant across the entire wavelength range considered, and except for the case shown in \figurename~\ref{fig:diff1}(a), the deviation remains stable over time.

Furthermore, it is worth noting that the QE measurements taken after exposure show greater fluctuation at the short ($\leq$ 350 nm) and long ($\geq$ 600 nm) wavelength limits. As highlighted in \cite{hamam}, this phenomenon can be attributed to two main factors. For longer wavelengths, it is primarily due to the strong dependence of the photocathode’s sensitivity on the uniformity of its surface. At shorter wavelengths, however, the uncertainty is further exacerbated by the physical properties of the borosilicate glass surrounding the photocathode, which significantly limits the transmission of ultraviolet radiation below 300 nm.

These observations allow us to formulate hypotheses about the behavior of the K-Cs-Sb photocathode under intense thermal and/or light stress, providing a rationale for the temporary degradation in QE. As noted in \cite{zihao}, prolonged exposure to temperatures above 77$^{\circ}$C may cause the release of cesium, which is not firmly bound within the crystalline structure of the photocathode. This results in an excess of cesium atoms that are "resting" on the internal surface of the lattice, making them easily releasable as temperature increases. A second contributing factor could be possible local alterations of the crystalline structure, which affects the uniformity of the photocathode's surface.


\section{Conclusions}

This study provides an evaluation of the effects of light and thermal stress on the performance of Sb-K-Cs bialkali-metal-coated PMTs. By investigating the behavior QE and dark current behavior, we demonstrate that both light and thermal exposure can significantly impact the sensitivity of PMTs. These effects are particularly relevant in specific wavelength ranges, especially between 350 and 550 nm, where PMTs show the highest variability in response to environmental stressors.

The results indicate that, in most cases, the degradation of QE due to light exposure and moderate thermal stresses is reversible, with recovery times varying based on the duration and intensity of exposure. However, more severe thermal stress, particularly at higher temperatures and with prolonged exposure times, can lead to irreversible damage to the PMTs, as evidenced by the permanent suppression of QE and persistently elevated levels of dark current. The observed degradations are primarily attributed to the evaporation of cesium, which occurs under high light and thermal conditions.

\clearpage

\end{document}